\documentstyle[12pt,aaspp4]{article}
\begin{document}
\tightenlines
%
%
\newcommand{\ea}{{et~al.}}
\newcommand{\IUE}{{\it IUE}}
\newcommand{\dmod}{$(m - M)_{0}$}
\newcommand{\logg}{$\log g$}
\newcommand{\logl}{$\log L$}
\newcommand{\lya}{\mbox{Ly$\alpha$}}
\newcommand{\lsun}{L$_{\sun}$}
\newcommand{\msun}{M$_{\sun}$}
\newcommand{\mv}{{M}$_{V}$ }
\newcommand{\rsun}{R$_{\sun}$}
\newcommand{\teff}{T$_{\rm eff}$}
\newcommand{\vsini}{$v \sin i$}
\newcommand{\rd}{Di\thinspace Stefano}
\newcommand{\re}{Einstein radius}
\newcommand{\er}{Einstein radius}
\newcommand{\ml}{microlensing}
\newcommand{\pl}{planet}
\newcommand{\mage}{magnification}
\newcommand{\lc}{light curve}
\newcommand{\ev}{event}
\newcommand{\ec}{encounter}
\newcommand{\pop}{population}
\newcommand{\mo}{monitoring}
\newcommand{\bi}{binary}
\newcommand{\bis}{binaries}
\newcommand{\ob}{observation}
\newcommand{\ps}{planetary system}
\newcommand{\sy}{system}
\newcommand{\pr}{program}
\newcommand{\res}{resonant}
\newcommand{\ov}{overlap}
\newcommand{\cs}{central star}
\newcommand{\inm}{innermost planet}
\newcommand{\dbn}{distribution}
\newcommand{\shdn}{short-duration}
\newcommand{\evd}{evidence}
\newcommand{\zres}{zone for resonant lensing}
\newcommand{\sep}{separation}
\newcommand{\dtn}{detection}
\newcommand{\bl}{blending} 
\newcommand{\mt}{monitoring teams} 
\newcommand{\fut}{follow-up teams} 
\newcommand{\ft}{follow-up teams} 
\newcommand{\fss}{finite-source-size} 
\newcommand{\fsse}{finite-source-size effects} 
\newcommand{\cc}{caustic crossing}
\newcommand{\Asun}{A$_{\sun}$}
\newcommand{\Ajup}{A$_J$}
\newcommand{\Anep}{A$_N$}
\newcommand{\Nsun}{N$_{\sun}$}
\newcommand{\goesto}{\longrightarrow}
\def\mr{multiple repetitions}
\def\op{orbital plane} \def\lp{lens plane}
\def\wo{wide orbit}
\def\otn{orientation} \def\vy{velocity} \def\vt{$v_t$}
\def\smn{simulation}
\def\em{Earth-mass}
\def\jm{Jupiter-mass}


\def\stacksymbols #1#2#3#4{\def\theguybelow{#2}
    \def\verticalposition{\lower#3pt}
    \def\spacingwithinsymbol{\baselineskip0pt\lineskip#4pt}
    \mathrel{\mathpalette\intermediary#1}}
\def\intermediary#1#2{\verticalposition\vbox{\spacingwithinsymbol
      \everycr={}\tabskip0pt
      \halign{$\mathsurround0pt#1\hfil##\hfil$\crcr#2\crcr
               \theguybelow\crcr}}}

\def\lapproxeq{\stacksymbols{<}{\sim}{2.5}{.2}}
\def\gapproxeq{\stacksymbols{>}{\sim}{3}{.5}}
\def\du{duration} \def\dca{distance of closest approach}
\def\stl{stellar-lens}
\vskip -.4 true in
\title{
A New Channel for the Detection of Planetary Systems Through Microlensing
}

\author{Rosanne \rd\altaffilmark{1},
Richard A. Scalzo\altaffilmark{2}}

\altaffiltext{1}{
Harvard-Smithsonian Center for Astrophysics,
60 Garden St., Cambridge, MA 02138; e-mail:  rdistefano@cfa.harvard.edu}

\altaffiltext{2}{Department of Physics, University of Chicago,
Chicago, IL 60637; e-mail:  rscalzo@rainbow.uchicago.edu}

\vspace{-0.15in}

\begin{abstract}

\vspace{-0.15in}
We propose and evaluate the feasibility of a new strategy to search for \pl s
via \ml\ \ob s.
This new strategy is designed to detect \pl s
in ``wide" orbits, i.e., with orbital separation, $a$,
greater than $\sim 1.5 R_E$.  
Planets in \wo s may provide the dominant channel for the  
discovery of \pl s via \ml, particularly low-mass (e.g., \em ) \pl s.
Because the ongoing \ml\ \ob s and extensions of them should be able to
discover \pl s in \wo s, we provide a foundation for the search 
through detailed calculations and simulations that quantify the expected
results and compare the relative benefits of various search strategies.  
If planetary systems similar to our own or to some of the known
extra-solar systems are common, then the predicted detection rates
of wide-orbit \ev  s are high, generally in the range $2-10\%$ of the
present detection rate
for apparently single \ev s by stars. The expected
 high rates should allow 
the \ml\ observing teams 
to either place significant limits
on the presence of \ps s in the Galactic Bulge, or begin to
probe the population in detail within the next few years.  
We also address the issues of (1) whether \pl s discovered via \ml\ are likely
to harbor life, (2) the feasibility of follow-up
\ob s to learn more about \pl\ microlenses, and (3) the contamination
due to stellar populations of any \ml\ signal due to low-mass MACHOs.
 
\end{abstract}
\vskip -.2 true in

\section{Introduction}

When planets are discovered through monitoring programs similar to those 
presently being carried out, the distance to the lens will typically be on the
order of kiloparsecs.
Although some spectral follow-up may be possible (\S 6), 
imaging, such as that planned
to probe possible companions to nearby stars (Stahl \& Sandler
1995; Bender \& Stebbins 1996; Labeyrie 1996; Angel \& Woolf 1997)
will not be possible in the near
future.
Nevertheless, planets discovered as microlenses can play an important role in
developing our understanding of planetary systems in our Galaxy and beyond.
The reason for this is precisely because microlensing observations probe vast
volumes of the Milky Way and other galaxies, such as the Magellanic Clouds and M31.  Thus,
microlensing provides a unique window for studying the statistics of planetary
systems (numbers and properties) and their dependence on the local stellar
environment.  Microlensing searches complement velocity-based searches in
another way as well.  Particularly for the wide systems studied here for the
first time, the microlensing searches can be effective for low-mass planets 
orbiting at low speeds.

The framework for the work to date on discovering planets through \ml\
was established by Mao \& Paczy\'nski (1991) and Gould \& Loeb (1992).
These authors found that when the separation, $a$, between the star
and \pl\ is close to the Einstein radius, $R_E,$ of the star 
($0.8 R_E  \lapproxeq a \lapproxeq 1.5 R_E$),
a significant
fraction of \ev s ($\sim 5-20\%$) in which the star serves as a lens
would be perturbed in a detectable way. 
This has been referred to as ``resonant" lensing,
both because the separation must be close to $R_E$ in order for
the signal to be detectable, and because the signal itself is
sharp and distinctive. 
\footnote{We will use the phrase ``\zres" to refer to the region in which
distinctive \ev s occur, whose characteristics tend to be related to
the presence of the caustic structure (i.e., \ev s 
of the type specifically studied by Gould \& Loeb [1992],
Bennett \& Rhie [1996],  
and  Wambsganss [1997]).  
The spatial extent of the \zres\ has been studied most
recently by Wambsganss (1997); our results do not rely heavily on
the exact position of its boundaries as a function of mass ratio.
}  
The detection strategy is to monitor light from an
ongoing event at frequent intervals in order to observe a short-lived
perturbation.
With the analysis of more than $200$ \ml\ \ev s reported to date,
it is not clear whether any planetary-lens \ev s have been discovered.
\footnote{
One candidate for a resonant planetary lensing event has been suggested
(Bennett {\it et al.} 1996).
There are, however, two reasons to be cautious
about the interpretation of this event.  First, the mass ratio derived from
the binary-lens fit is $\sim 0.043$, which is large enough to be consistent
with lensing by a binary stellar system.  Second, the degeneracy of the
physical solution has not yet been worked out.  The degeneracy may be of two
types:  (a)\ Other binary solutions with values of $q$ differing from the one
for this fit by as much as an order of magnitude may prove to be equally good
fits (Di Stefano \& Perna 1996); this needs to be systematically checked.
(b)\ Other physical effects may prove to be important, 
such as 
finite source size
effects (Witt \& Mao 1994; Witt 1995), finite lens-size effects
(Bromley 1996), and 
blending (\rd\ \& Esin 1995, Kamionkowski 1995).
(In fact blending has been  
a feature of every other binary-lens event; Udalski {\it et al.} 1994,
 Alard, Mao, and Guibert 1995, Alcock {\it et al.} 1997a.)
}
Indeed, the fraction of \ml\ \ev s to date in which evidence
of a planet in a resonant orbit is detectable is 
apparently less than $\sim 1\%$.
This is not necessarily inconsistent with the original predictions, however,
because planets with the appropriate distance from the central star may not
be common.  In addition, finite-source-size effects can wash out the signature
of the short-time-scale perturbations characteristic of planetary mass ratios,
decreasing the number of detectable \ev s (Bennett \& Rhie 1996).

The purpose of this paper is to develop a complementary framework for the
discovery of \pl s by \ml . The basic idea is that planets located more than
$\sim 1.5 R_E$ from the central star can give rise to \ml\ \ev s that reveal
evidence of their presence.  We will say that planetary orbits in which the
separation between the star and planet is larger than 1.5 $R_E$ are ``wide".
Concrete definitions as well as our expectations for the possible
existence of a population of
wide-planetary-orbit lenses are discussed in \S 2. Because the ongoing
observational programs have already developed the tools needed to begin a systematic
search for \pl s in \wo s, the rest of the paper is designed to give
a detailed view of what observers can expect to see and to compare
different observational strategies.
  In \S 3 we
focus on the character of the events expected when the lens is a wide
planetary system.  In \S 4, we present the results of Monte Carlo
simulations of lensing by several model planetary systems, including
our own solar system, to test basic
features of detectability. In \S 5 we study the link between the search
for evidence of \pl s in wide orbits and the search for extraterrestrial
life.
Blending may play an important role in allowing us to learn more about
\ps s, including the mass of a \pl\ and the spectral type of the
central star; blending is the topic of \S 6. When the size of the source is
comparable to or, to some extent, larger than the Einstein ring of the
\pl , the probability of detecting \pl s in wide orbits can {\it increase}.
This is in contrast
to lensing by \pl s located in the \zres , where there is a tendency
for detection probabilities to decrease as the ratio between the source radius and
the \re\ increases. 
\footnote{
A more precise statement is that for a \pl\ of a given mass,
there is a range of source sizes for which \fsse\ 
influence the rate of detection,
but do  not prevent the detection of \pl-lens \ev s.
Within this range, \fsse\ tend to increase the rate of detection of
\pl-lens \ev s when the \pl\ is in a \wo, and to decrease
the detection rate for an identical \pl\ located in the \zres . 
The net effect is to increase the relative rate of detectable wide-orbit
lensing \ev s.} 
Thus, \pl s in \wo s may provide the primary channel for the
discovery of low-mass (e.g., \em )  \pl s.
Finite-source-size effects are the subject of \S 7.  
In addition to including multiple \pl s, \ps s may be more complex,
containing moons bound to \pl s, and also
 belts or clouds of asteroids and comets.
Because the detectability of these features tends to be limited by \fsse ,
we discuss them in \S 8. 
\S 9 is devoted to a discussion of the
detection strategies that will improve our ability to
discover \pl s in wide orbits. In \S 10 
we address the likely near- and long-term
results of implementing those strategies.
Finally, in \S 11 we summarize our results.

\section{Wide Planetary Systems}

\subsection{Definition}

We will say that a planet is in a ``wide" orbit if its distance from the central
star is large enough that the isomagnification contour associated with
$A = 1.34$ is comprised of two separated, closed curves, one centered near the
star and the other centered near the planet.  For mass ratios ranging from
$10^{-5}$ to $10^{-3}$, the critical orbital radius, $a_w$, beyond which an
orbit becomes wide is roughly equal to $1.5 R_E$.  It is also useful to
define an inner radius, $a_c$.  For $a < a_c$ the fraction of all light curves
displaying deviations from the standard point-lens form is small (generally
less than 1\%).  For $a_c < a < a_w$ a larger fraction of light curves
exhibit perturbations, including caustic crossings.  The value of $a_c$ is
approximately equal to $0.8 R_E$ for values of $q$ appropriate to planets
(Gould \& Loeb 1992).

In Figure 1 we consider planetary lenses in close, resonant, and
wide orbits; in each case the mass ratio between the planet and the star
is $0.001,$ comparable to that between Jupiter and the Sun.
Thousands of \ev s were generated in which the track of the source
passed behind each lens; in this way we could sample the results expected when
the system serves as a lens. In the plots of distributions of
event durations, the most striking feature is the
appearance of a peak corresponding to short-duration events
when the orbital separation
is wide (right-most panel). 
Indeed, 3\% of the events for $a = 3 R_E$
have durations less than 2 days.  
Lensing by stellar systems produces such a distinct peak of \shdn\ \ev s
only when there are wide-orbit \ps s 
in which \pl s serve as isolated lenses. 
Other features illustrated by the figure include the following.
When a \pl\ is in a close orbit, the
probability of detecting its presence via \ml\ is small. At ``resonant"
separations, the fraction of \ev s revealing evidence of the caustic structure
associated with the presence of the \pl\ is significant. While the fraction
of such \ev s decreases as the orbital separation increases toward the
``wide" regime, a significant
fraction of lensing \ev s in the wide regime
are (1) the short-duration events mentioned above, 
which are absent in the other cases, and/or
(2) multiple-peak events, in which the separation between two of the
peaks is so large, and the magnification between them falls so low,
that they are best thought of as repeating \ev s.

\begin{figure}
\vspace{-1 true in}
\plotone{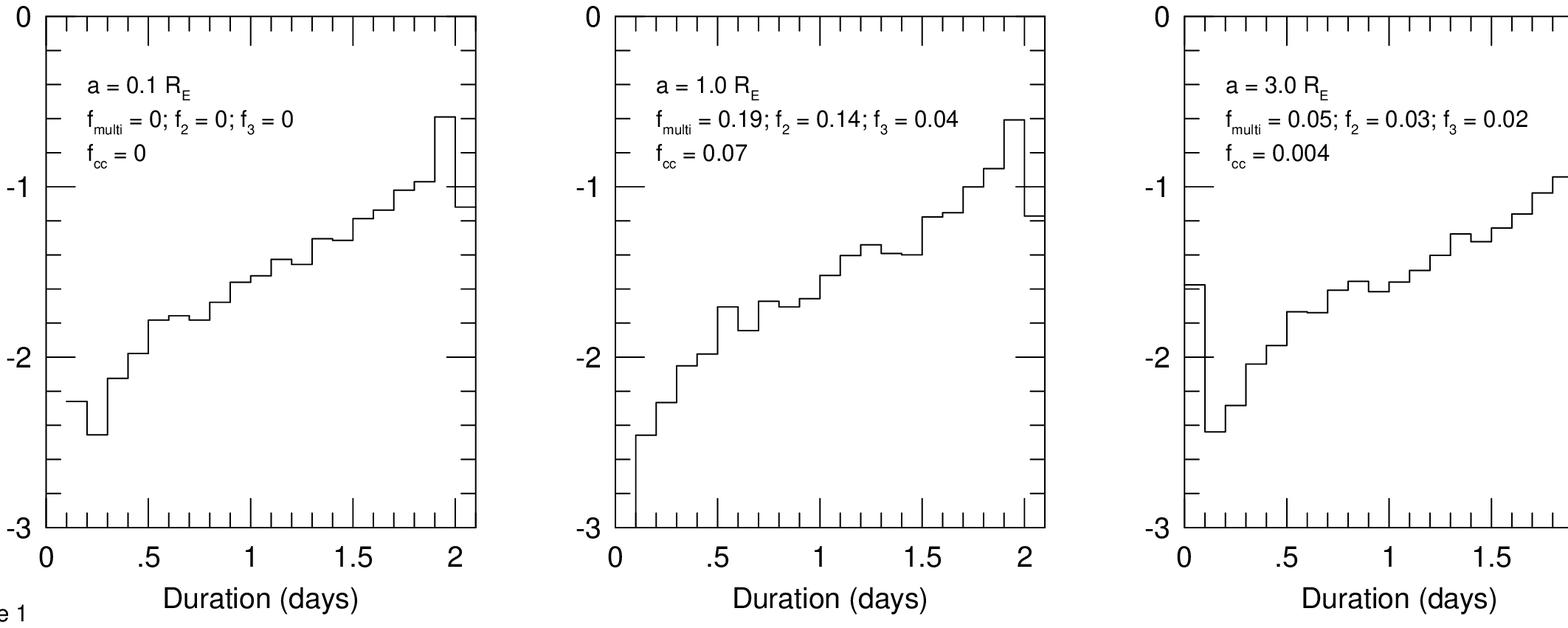}
\vspace {-4 true in}
\caption{Plotted is the distribution of \ev\ durations for 3 microlenses.
Each lens has a \pl\ of mass $m_i$ in
orbit with a star of mass $M$; in each case we chose $q=m_i/M$ to be
$10^{-3}$, the mass ratio between Jupiter and the Sun. The lenses differed
from each other only in the value of the orbital separation, $a$.
For each lens, several thousand source tracks were randomly generated,
and the \lc\ was computed for each \ev .
$f_{cc}$ represents
the fraction of all \ev s which exhibited caustic crossings;
$f_{multi}$ ($f_2$, $f_3$) represents
the fraction of all \ev s which exhibited multiple ($2$ ,$3$) peaks.
}
\end{figure}

\subsection{Extending Planetary Searches Beyond the Resonant Case}

The excitement about resonant lensing by planets was fueled in part by a
wonderful coincidence. Gould \& Loeb (1992) noted  that,
if a system identical to our solar system happened
to be located halfway between our position and the center of the Bulge, and if
the system were viewed face-on, the separation of the planet corresponding to
Jupiter from the system's star would be very close to the value of $R_E$
associated with the mass of this star.  That is, Jupiter would be in the 
\zres . There are,
however, two reasons to be cautious about using this example to limit the
search for planets to those in resonant orbits.  First, we do not know that
the spatial relationship between the Sun and Jupiter is an example of a
universal property of planetary systems. 
Mindful of this, Bennett \& Rhie (1996) have
constructed a ``power-of-2'' planetary model, in which the separation between
each planet and the central star increases by a factor of two for each
successive planet.  In such a model, most planetary systems can be expected to
contain one planet in a resonant orbit.

A second reason to avoid limiting the microlensing searches to resonant
planets is that the value of the Einstein radius depends not only on the
stellar mass, but also on the relative positions of both source and lens to
the observer.
\begin{equation}
R_E=\Bigg[{{4\, M\, G\, D_S\,\, x\, (x-1)}\over{c^2}}\Bigg]^{{1}\over{2}}. 
\end{equation}
where $M$ is the mass of the lens, $D_S$ is the distance from 
the observer to the lensed source, and $x=D_L/D_S,$ with $D_L$ 
representing the distance from 
the observer to the lens.

In Figure 2 we consider planets with orbital separation  
$a=2\, a_w = 3\, R_E$. For these systems we show
the relationship between the stellar mass, $M$, and $x$, when the
orbital period is fixed.
We have
assumed that the orbital plane is the same as the lens plane, 
and show those values of $x$ for which both lens
and lensed source are located
in the source galaxy (the Bulge, the Magellanic Clouds, or M31). If the
stellar mass is in the range from $0.1$ to $10\, M_\odot,$ the orbital 
periods of \pl s in \wo s range from a few years to a few hundred
years, with larger values more typical for more distant galaxies.

\begin{figure}
\vspace{-1 true in}
\plotone{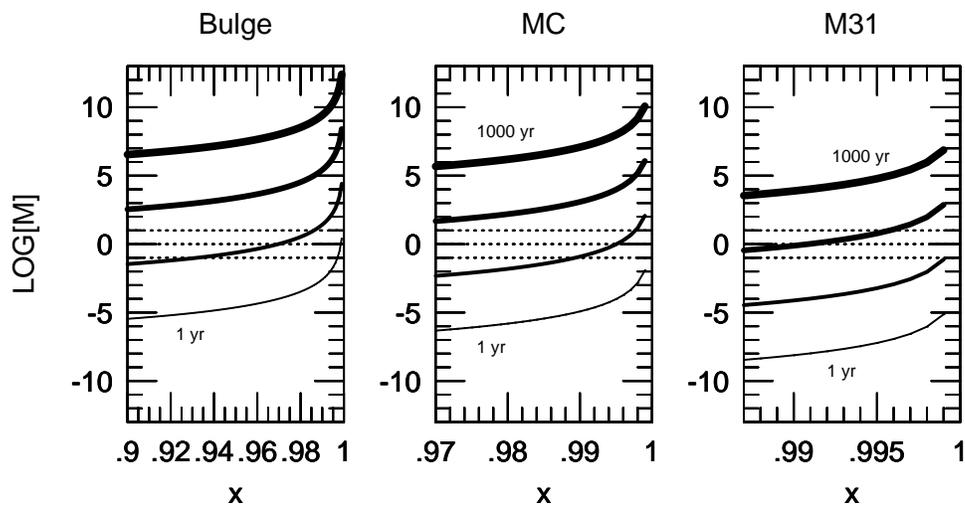}
\vspace{-4 true in}
\caption{
Each curve corresponds to a fixed value
of the orbital period; shown are
curves for 1 yr, 10 yrs, 100 yrs, 1000 yrs.
Each curve illustrates the relationship between the stellar mass, $M,$ 
and $x$ for a 
\pl\ in a \wo, with $a=3\, R_E$. In each panel the range of the
variable $x$ corresponds to lenses located within the 
same galaxy (Bulge, Magellanic Clouds, M31 [left to right])
as the lensed stars. 
Note that the range of masses and orbital periods best explored via
wide lensing events is different for different source galaxies.
}

\end{figure}

In addition, the projected value of the separation depends on the
angle, $\alpha$, between the normal to the orbital plane and our line of
sight.
Consider an orbit with $a > a_w$. Let $\alpha$ represent the angle between
the normal to the plane of the orbit and the normal to the lens plane.
For $\alpha < \alpha_{min}$, where
\begin{equation}
\alpha_{min} = \cos^{-1}(a_w/a),
\end{equation}
the projected value of the orbital separation is always greater than $a_w$.
If $\alpha > \alpha_{min}$, then for a circular orbit, the fraction of time
for which the projected separation is greater than $a_w$ (so that the planet
can be viewed as wide) is
\begin{equation}
f = \frac{2}{\pi}
    \cos^{-1} \left(
            \frac{\sqrt{\cos^2 \alpha_{min} - \cos^2 \alpha}}{\sin \alpha}
            \right).  \label{frac-crw}
\end{equation}

\noindent
If the orbit is eccentric, then the separation between the planet and star will
be wide for an even greater fraction of the time.

Thus, even if planetary systems had uniform properties, the fact that
they are located at different spatial positions and are tilted at different
angles relative to our line of sight would still not favor resonant orbits over
others.  Bennett \& Rhie's power-of-2 model, or something like it, is
necessary to ensure that most planetary systems have one planet in the zone
$a_c < a < a_w$ associated with resonant lensing.  This is because
$a_w/a_c \approx 2$; thus, if the $i^{th}$ planet has separation $a_i$ from
the central star, and the power-of-2 system is inclined so that
$a_i \cos \alpha \approx a_c$, we have $a_{i+1} \cos \alpha \approx a_w.$

The known planetary and brown-dwarf binary systems are considered in Table 1,
together with a ``power of 2'' model and a ``power of 3'' model.
The known systems included in the table were selected from the
Encyclopedia of Extrasolar Planets 
(accessible from www.obspm.fr/darc/planets/encycl.html). Two
criteria were used: (1) if the \ps\ were placed in the Bulge and viewed face-on,
the planet (or brown dwarf) listed had to be in an orbit that
would either be wide or located in the \zres , (2) the existence of the \pl\
or brown dwarf 
needed to be listed as ``confirmed".
 In the power-of-2 and power-of-3
model, the first planet listed is the innermost \pl\ that 
would be in a \wo, were the \ps\  to be placed in the Bulge and viewed face-on.
We note that the next 
\pl\ inward would have a good chance of being viewed in the \zres .
Indeed for both of these theoretical models, there is some
ambiguity associated with the arbitrarily-chosen position of the
closest \pl\ to the central star. Because of this, we carried out
a set of calculations in which we averaged over the position of the innermost
\pl .
We found that, if we truncate the
radius of the \ps\ at $\sim 10^{17}$ cm, 
there are on average $\sim 9-10$ \pl s in \wo s for
every \pl\ located in the \zres.    
To derive the numbers shown in the table, we averaged over inclination angle
$\alpha$, and used equations (2) and (3) 
to determine the fraction of time any given planet would be in a wide orbit.
As above,
we have assumed that the orbits are circular.

The considerations in this section show that planets in wide orbits exist,
and may be common.
In fact the number of planets in wide orbits may be as much as
an order of magnitude larger than the number of planets in resonant
orbits. The relative contribution of these
two classes of planet lenses to the detection rate depends on
our ability to identify the associated events.

\tightenlines
\begin{deluxetable}{llrlll}
\label{visibility}
\scriptsize
\tablecaption{Planets in wide and resonant orbits in known and model systems;
\hfil\break $D_s = 10$ kpc, $x = 0.9$.}
\tablehead{\colhead{Planetary system} & {Planet} &
   \colhead{$a$ \tablenotemark{(1)} } &
   \colhead{$\%_{close}$ \tablenotemark{(2)} } &
   \colhead{$\%_{res}$ \tablenotemark{(3)} } &
   \colhead{$\%_{wide}$ \tablenotemark{(4)} }} 
\startdata
{\bf Solar system:} 
& Jupiter & 5.2 & \, 5.9  & 21.0 & 73.1 \\
& Saturn  & 9.5 & \, 1.7  & \, 4.6  & 93.7 \\
& Uranus  & 19.2 &\, 0.4  & \, 1.0  & 98.6 \\
& Neptune & 30.1 &\, 0.2  & \, 0.4  & 99.4 \\
& Pluto   & 39.8 &\, 0.1  & \, 0.2  & 99.7 \\
\hline
{\bf Known extrasolar systems:} 
& 55 Cnc       & 4.0 & 11.0 & 89.0 & \,\, 0.0 \\
& HD 29587     & 2.5 & 37.0 & 63.0 & \,\, 0.0 \\
& PSR B1620-26 & 38 & \,\, 0.1 & \,\, 0.3 & 99.6 \\
& Gl 229b &    40 & \,\, 0.1 & \,\, 0.2 &  99.7 \\
\hline
{\bf Theoretical power-of-2 system:}
& 1 & 4.8 & \,\, 7.0 & 27.0 & 66.0 \\
& 2 & 9.6 & \,\, 1.6 & \,\, 4.5 & 93.9 \\
& 3 & 19.2 & \,\, 0.4 & \,\, 1.0 & 98.6 \\
\hline
{\bf Theoretical power-of-3 system:}
& 1 & 4.8 & \,\, 7.0 & 27.0 & 66.0 \\
& 2 & 14.4 & \,\, 0.7 & \,\, 1.9 & 97.4 \\
& 3 & 43.2 & \,\, 0.1 & \,\, 0.2 & 99.7 \\
\hline
\enddata
\tablenotetext{(1)}{Separation of the planet from the central star
in astronomical units (AU). We
consider circular orbits. 
See the text for the criteria used to select the \pl s listed here.} 
\tablenotetext{(2)}{Percentage of the time during which the planet is
``close'' ($a < 0.8 R_E$), averaged over all inclinations.}
\tablenotetext{(3)}{Percentage of the time during which the planet is
``resonant'' ($ 0.8 R_E < a < 1.5 R_E$), averaged over all inclinations.}
\tablenotetext{(4)}{Percentage of the time during which the planet is
``wide'' ($a > 1.5 R_E$), averaged over all inclinations.
In the power-of-2 and power-of-3 models,
all planets beyond the third had a probability greater than $99\%$
of being located more than $1.5\, R_E$ from the central star.}
\end{deluxetable}

\section{The Detection of Wide Planets}

There are two distinctive signatures of the presence of wide planets in the
stellar systems that serve as lenses.  The first is a significant number of
short-duration events in the data sets.  Although relatively few individual
short-time-scale \ev s may be unambiguously identifiable as being due to
planets, the statistics associated with the presence or absence of such \ev s
provide a strong diagnostic for properties of the population of \ps s.
The second distinctive signature is a
small number of repeating events. 
Repeating events are good
diagnostics for individual \ps s, yielding a value for the mass ratio and the
orbital separation (expressed in units of $R_E$) as projected onto the plane
of the sky.

For both isolated and repeating events,
if the central star is luminous and provides a significant
fraction of the light received along the line of sight to the lensed source,
then spectroscopic studies, conducted during the \ev\ and also at baseline,
can determine the central star's spectral type and may allow us to derive the
mass of the planet (\S 6). Another diagnostic of the \pl-lens mass for
both isolated and repeating \ev s is provided by finite-source-size effects,
which can,
in addition to providing information about the mass of the planet-lens, 
also help to increase the detection rate of low-mass \pl s in wide orbits (\S 7).

\subsection{Notation: Encounters, Events, and Repetitions}

The \ps s we consider consist of a central star 
and $N$ \pl s moving in wide, circular orbits.
Let the index $i$ label the central star ($i = 0$) 
and the planets ($i=1,N$). 
The characteristics of the \ps\ are
specified by the mass of the central star, $M=m_0,$ 
 the planetary masses, $m_i$, the associated mass ratios, $q_i=m_i/M$,
and the orbital radii, $a_i$, of the \pl s.   
For the sake of convenience, we will sometimes drop the index
for the star; 
unless otherwise specified, $M$ and $R_E$ refer to the mass and
Einstein radius of the star. 

\subsubsection{Width of the Lensing Region}

Let $A_{min}$ be  
the minimum value of the peak \mage\ for which a perturbation
is reliably ascribed to \ml . 
Each value of $A_{min}$ is associated with
a specific value of $n,$ the number of Einstein radii equal to
the minimum necessary distance of closest approach.  
If $A_{min}$ is $1.58$ ($1.34$, $1.06$, $1.02$), then 
$n=0.76$ ($1$, $2$, $3$).
Note that $n$ need not be an integer.

For any given lens, the ``width" of the lensing region is defined to be
\begin{equation}
w_i = n\, R_{E,i}.
\end{equation}
Thus, if $A_{min}=1.34,$
the width associated with each mass is simply equal to its Einstein
radius. With this definition of the width, the 
probability of an \ec\ with a particular lens
can be taken to be $2\, w_i= 2\, n\, R_{E,i}.$

\subsubsection{Encounters and Events}

When a \ps\ with several \pl s in \wo s serves as a lens, the track of the
source may pass through the region of influence of several lenses. 
When the 
source track passes within $w_i$ of lens $i$, we will say that an ``\ec"
is underway. We will use the word ``\ev" to refer to a source track 
(and the associated \lc) that experiences one or more \ec s.      

\subsubsection{Repeating, Isolated, and Overlap Events}

When more than one \ec\ occurs, we will dub the \ev\ a 
``repeating" \ev ; \ev s with just a single \ec\ will be called
``isolated".   

The relatively short time duration of \ec s in which a \pl\ serves
as a lens means that frequent \mo\
is needed 
to detect them.
If a stellar-lens \ec\ takes 100 days, a Jupiter-mass lens \ec\ will last
$\sim 3.3$ days, and an Earth-lens \ec\ will last $\sim 8$ hours.
(Finite-source-size effects can, however, increase the duration of 
detectable \ec s.
We consider them in \S 7.)
Thus, without frequent \mo, we may miss the occurrence of an isolated \pl-lens \ev. 
If one component of a repeating \ev\ involves the central star, then we 
may see that component, and miss the \pl-lens component. 
To increase our chances of discovering repeating \ev s, 
we can use the
fact that the \ec\ due
to the star is likely to occur first roughly half the time, 
and continue frequent \mo\ of lensing \lc s, even after they have
returned to baseline.
The calculations we describe in \S 4 indicate  that
$\sim 50-75\%$ of the cases in which a \pl\ will subsequently serve as
as a lens can be discovered if frequent \mo\ continues for $100$ days.
Similarly, moons revolving about \pl s could be discovered with frequent \mo\
lasting for a time interval on the order of days, subsequent to a \shdn\
\ec\ which might be due to lensing by a \pl\ (\S 8).

\begin{figure}
\vspace{-2 true in}
\plotone{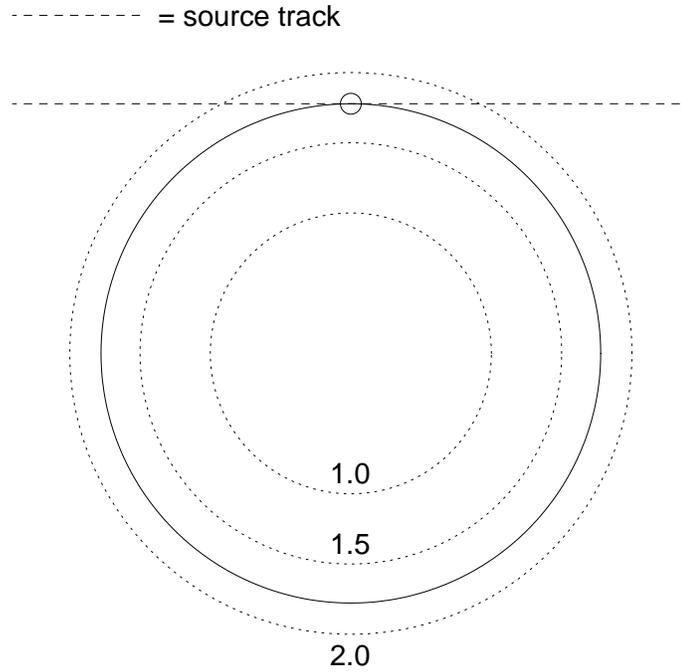}
\vspace{-2 true in}
\caption{The innermost \pl\ of the power-of-2 system.  
The dotted circles represent
the 1.0--, 1.5--, and 2.0--$R_E$ rings for the central star.  The large solid
circle is the orbit (viewed face-on)
of the innermost wide planet, at a separation of 4.8 AU.  
If the central star is of solar mass, then the radius of the small solid
circle on the orbit is roughly equal to the Einstein radius of a 10-Jupiter-mass
object.
When $w_0 = R_E = w_{0,0},$  
there is a range of angles along which a source track can pass
through the Einstein ring of the planet without
also encountering the star. When $w_0 =  2\, R_E = 2\, w_{0,0},$ 
any encounter with the
planet also results in an encounter with the star.    
}
\end{figure}

The ideal case is that in which the perturbation due to lensing by one mass
has not yet ceased before lensing by the second mass begins to have a noticeable
effect. 
We will refer to such \ev s as ``overlap" \ev s.
Overlap \ev s can occur when the separation, $a,$ between the lenses is less
than $w_1+w_2.$ As $a$ decreases below $w_1+w_2,$ many \ev s which
involve \ec s with both lenses will give rise to \lc s in which the 
\mage\ does not fall below baseline between \ec s. These \lc s are similar
to others exhibiting repetitions, except that the perturbation due to
the first \ec\ is still noticeable when the second \ec\ begins.
We are almost guaranteed
to discover $1/2$ of all repeating ``overlap" \ev s, even with no or
little change in detection strategy. 
As the value of $a$ decreases further, a larger fraction of the \ec s with
the smaller lens are part of \ev s in which the larger lens is
also encountered; for $a < w_1-w_2,$ this is true of all \ec s involving the
smaller lens. Most of the associated \lc s will exhibit a single connected
perturbation, with the structure in the wings largely
determined by the most massive lens, and the structure near the peak
largely determined by the less massive lens.
Although such \lc s may appear to be highly anomalous,
we should be able to fit them to the appropriate lensing model.
Since the \lc\ perturbations are long-lived (relative to \ec s with
an isolated \pl ),
we might be able to detect and identify $100\%$ of such overlap \ev s.
(See Figure 3.)  
      
In general we expect at most $N+1$ \ec s and $N$ repetitions.
In principle, if the source travels slowly and the \pl s orbit
rapidly, a single \pl\ could give rise to more than one \ec .
In practice this is extremely rare (\S 3.4.6). Repetitions, however,
are not rare. As we will see, if most stars have \ps s similar to
our own, a data set containing the number of \ev s already observed 
along the direction to the Bulge should contain one or more \ev s
in which a repetition is due to lensing by a \wo\ \ps. As we will also
discuss, the detection of such \ev s can be optimized by frequent
sampling with sensitive photometry. 

\subsection{Normalization of Event Rates}  

We will be interested in the rate at which detectable \ev s occur,
 and how that rate is influenced by changes in the detection strategy.
The Einstein radius of the 
central star provides a convenient normalization.
The rate of \ec s in which the central star serves as a lens,
with the magnification due to the central star achieving a peak value 
of at least $1.34$ ($A_{min} = 1.34$), is proportional to 
$2\, R_E.$ Because the lensing width of the central star (when $A_{min} = 1.34$) 
will play an important role, we define $w_{0,0}$ to be equal to $R_E$.  
If the photometric sensitivity is such that $A_{min}$ can be smaller than
$1.34$ (i.e., $n>1$), so that the effective width of \ev s involving the central
star is larger ($w_0 > w_{0,0}$), the rate of all detected \ev s,
including those due to the central star,
will increase. 
Nevertheless, 
 $2\, w_{0,0}\, $ is a convenient normalization constant,
because it allows us to compare the rates we compute to
 the presently-measured rate of \ev s.
This is as follows.
Along the direction to the
Bulge, \ev s 
 are being discovered by the MACHO team at a rate of
roughly $50$ per year (See \S 9 for the distribution of observed \du s.) 
The present detection criteria used by the MACHO team
are strict: $A_{min}=1.58$, corresponding to $n$ (in Eq.\ 4)
being set equal to $\sim 0.76.$   In addition, other teams,
surveying different fields with somewhat different strategies,
are also surveying the Bulge. For example, the OGLE team,
which in its first incarnation discovered $\sim 12$ Bulge \ev s,
has recently
brought a new telescope on-line (Udalski, Kubiak \& Szymanski 
1997; Paczy\'nski {\it et al.} 1997). 
It is therefore
reasonable to assume $75-100$ \ev s of the type
we use for our normalization per year along the direction to the 
Bulge. Thus, because we use $2\, w_{0,0}$ to normalize our results, when
we find  
that a particular detection
strategy leads to a rate of detectable 
\ev s of a certain type (e.g., \ev s with $2$ repetitions)
equal to $p\, \%,$ this means that between $0.75\, p$ and $p$ such \ev s
could be discovered per year along the direction to the Bulge.

\subsection{Isolated Events}

When a wide planetary system serves as a lens,
the most common type of event
is one in which the track of the source passes through the Einstein ring of
a single mass.  Because the rate of \ec s associated with a given lens scales
as the square root of its mass, \ec s in which the star serves as a lens are
the most frequent.  
Using the normalization described above, the rate of lensing due to \pl s 
is approximately given by 
\begin{equation}
P_1 \sim \sum_{i=1}^{N} n\, \sqrt{q_i},
\end{equation}
Because isolated \ev s are generally $1-2$ orders of
magnitude more common than repeating \ev s, $P_1$ generally provides
a good estimate of the rate of isolated events involving \pl s.
In the next subsection, we will focus on the rates of repeating \ev s,
which should be subtracted from the above expression to make it exactly
correspond to the rate of isolated \ev s.
The large majority of short-duration isolated events will be
indistinguishable from the standard point-mass-lens light curve.  A small
number will exhibit caustic crossings.  

The time duration of isolated
planetary-lens events, relative to the time duration of stellar-lens events,
also scales as $\sqrt{q_i}$.
The implication is that 
a distribution of events due to lensing
by stars with wide-orbit planets
is necessarily accompanied by a distribution
of shorter-duration events.  The fraction of events in the latter
distribution is proportional to the average value of $\sqrt{q}$, and the position of
the peak or peaks also provides a measure of the mass ratios typical of
planetary systems.

\subsubsection{Identifying Peaks In The Distribution of Short Events}

A peak corresponding to \shdn\ \ec s in the distribution
of \ec\ durations
should be unambiguously identifiable
as being due to the presence of lenses with small masses.
We argue as follows that such a peak is unlikely
to be hidden by the short-duration tail of \ec s due to lensing by
stars.
Three circumstances can make a stellar-lens \ec\ have a short
measured \du .

(1) The relative velocity $v_t$ an observer would measure between the
source and the lens may be exceptionally large.  Since, however, the
distribution of transverse velocities for planet-lens and stellar-lens
\ec s should be roughly the same (see Eq. \ref{planetvel}),
we should be able to disentangle velocity effects, at least
for statistical samples of \ec s.

(2)
The track of the source may just graze the lens' Einstein ring.  In this
case, the peak value of the magnification allows us to measure
$b,$ 
the distance of closest approach [which we will also call $D_{min}$], 
and to determine
that, although short, the observed \ec\ was nevertheless
due to the presence of a lens with a larger Einstein ring
than expected for a planet.
These corrections are implemented by the observing teams.

(3) Light from the lensed source could be blended with light
from other sources along the line of sight. If,
however, the effects of blending are dramatic enough to shorten the
\du\ of an \ev\ significantly, then we should
be able to detect evidence for the blending and to
subtract its effects from the event before comparing its \du\ with
that of other \ev s.
Because blending can provide a valuable tool for the study of planet-lenses,
we discuss it in more detail in \S 6.

\subsection{Repeating Events}

\subsubsection{One Repetition}

When the source track passes close
to one wide planet, and also within the
region of influence of the central star and/or that of other wide
planets, the \ev\ will appear
to repeat. Such repeating events were studied by Di\thinspace Stefano \&
Mao (1996) in the context of wide binaries--i.e., binaries in which
the separation between the components is large enough that
the $A = 1.34$ isomagnification contour consists of two
disjoint closed curves. 

Let $w_1$ ($w_2$) represent the width associated with lensing by the more 
(less) massive 
lens. 
We assume that the separation between the two lenses is wide--i.e., the
isomagnification contours associated with $A=1.34$ are distinct. As we
discussed in the previous section, this
generally means that if one lens is the central star and the other is
a \pl, $a > 1.5\, R_E.$ \footnote{Dynamical
stability of the \ps\ implies that two planets
will generally be separated from each other  
by significantly more than $1.5$ times the Einstein radius of the more massive \pl .}                      
The widths, $w_1$ and $w_2$, however can be smaller or larger than
the orbital separation, since their values are tied to issues of
photometric sensitivity. 

With $w_{0,0}$ defined as above, the detection rate
for repetitions involving lenses $1$ and $2$ is given by:
\begin{equation} 
{\cal R}_{1,2} = {{2}\over {\pi}} \, {{1}\over {2\, w_{0,0}}} 
\Bigg\{ \theta_{max} (w_1 + w_2) + \theta_{min} (w_2 - w_1) 
        + a\, \Bigg[ cos(\theta_{max})-cos(\theta_{min}) \Bigg] \Bigg\}
\end{equation}
In this expression, the values of $\theta_{max}$ and $\theta_{min}$ depend
on the value of the orbital separation, $a,$ as compared to the 
widths, $w_1$ and $w_2.$ 

\noindent For $a > w_1+ w_2 $,
\begin{equation} 
\theta_{max}=sin^{-1}({{w_1+w_2}\over{a}});
\, \theta_{min}=sin^{-1}({{w_1-w_2}\over{a}}).
\end{equation}

\noindent For $w_i+w_2 > a > w_1-w_2$,
\begin{equation} 
\theta_{max}={{\pi}\over{2}}; 
\, \theta_{min}=sin^{-1}({{w_1-w_2}\over{a}}).
\end{equation}

\noindent For $a < w_1-w_2 $,      
\begin{equation}  
\theta_{max}=\theta_{min}={{\pi}\over{2}}. 
\end{equation}

The limit $ w_1+ w_2 << a $  
corresponds to the case in which the separations
are extremely wide. In this case,
\begin{equation}
{\cal R}_{1,2} \cong {{2}\over{\pi}} {{w_1\, w_2}\over{a\, w_{0,0}}}
= n^2\,
\Bigg[ {{2}\over{\pi}} {{R_{E,1}\, R_{E,2}}\over{a\, w_{0,0}}}\Bigg]. 
\end{equation}
The rate of repetitions is inversely proportional to the 
separation. Another key feature of this expression is the 
quadratic dependence of the width of the lensing region,
which is directly related to the photometric sensitivity.

The second extreme limit corresponds to the case when the separation
between the lenses is smaller than $w_1.$ In this case,
\begin{equation}
{\cal R}_{1,2} \cong {{w_2}\over{w_{0,0}}} = n\, 
\Bigg[{{R_{E,2}}\over{w_{0,0}}}\Bigg]. 
\end{equation}
That is, the rate of repeats involving lenses 1 and 2 is the same as
the rate of \ev s involving 2; all source tracks passing   
through the lensing region associated with 2 necessarily
pass through the lensing region associated with 1.

\subsubsection{Estimates} 

For the purposes of \ml , the most important difference between a
stellar binary and a \ps\ is that the \ps\ may contain several planets.
A naive generalization of Eq.\ 10 leads to the following
expression for the average probability of repeating events.
\begin{equation}
P_2 = \frac{2 \, n^2}{\pi T_{sys}} \int_0^{T_{sys}}
   \left[ \sum_{i=1}^{N}
          \frac{\sqrt{q_i}}{|\vec{a}_i(t)|} +
          \sum_{i=1}^{N-1} \sum_{j>i}
             \frac{\sqrt{q_i q_j}}{|\vec{a}_i(t) - \vec{a}_j(t)|} \right]
   \, dt,  \label{naive-approx}
\end{equation}
where $T_{sys}$ is the time taken for the configuration of the \ps\
to approximately repeat, and $\vec{a}_i(t)$ is the position vector
of the $i\,$th \pl , expressed in units of the stellar \er .
Note that we are considering the regime $a >> w_1 + w_2;$ thus 
the rate of overlap \ev s is not included in these calculations.
 
To develop a feeling for the numbers, we will consider a simple
model, a
power-of-$k$ model, in which the plane of the
\ps\ is coincident with the lens plane and each planet has the same mass
ratio $q$ with the central star.
Using
\( \int_0^{T_{sys}} dt |\vec{a}_j(t)-\vec{a}_i(t)|^{-1}
\sim (a_j^2 + a_i^2)^{-1/2},\) allows us to simplfy 
equation (\ref{naive-approx}) in a way that does not overestimate 
the rate of wide-orbit lensing \ev s. 
\begin{equation}
P_2 = \frac{2\, n^2 \sqrt{q}}{\pi\, a_1}
   \left[ \sum_{i=0}^{N-1} \frac{1}{k^i} +
          \sum_{i=0}^{N-2} \sum_{j>i}
             \frac{\sqrt{q}}{k^i \sqrt{k^{2(j-i)} + 1}} \right].
\end{equation}
\noindent where $a_1$ is the distance between the star and the first
wide planet.
{For the power-of-2 model the value of the first term in
the above equation is approximately $2$, and that of the second term is
roughly $1.86\sqrt{q}.$
The rate of repeating \ev s
in which the star and one \pl\ each serve as a lens is 
$\Big[(4\, n^2 \sqrt{q})/{(\pi\, a_1)}\Big]$. If $q=0.001$ and $a_1=2,$ this
becomes $0.02\, n^2.$
\footnote{We note that, if $n$ approaches $2,$
then a \pl\  in an  orbit with $a=2$ would be detected through overlap \ev s;
$a$ would need to be larger for the approximation $a >> w_1+w_2$       
to apply. 
} 
The observed ratio between the rate of repeats involving 
both the star and one \pl , to the rate of stellar-lens \ec s is $0.02\, n.$
The ratio of such repeats to isolated \pl-lens \ec s is roughly 
$\Big[({2\, n})/({\pi\, a\, N})\Big],$ 
where $N$ is the number of \pl s in \wo s.
Note that the contribution of repeating \ev s saturates at moderate
values of $N$, while the rate of isolated \shdn\ \ev s is 
proportional to $N$. 
The ratio of repeating \ev s in which both \ec s are due to \pl\ lenses, 
to those in which one component is due to the central star, is suppressed 
by a factor ${\cal O}(\sqrt{q})$.

\subsubsection{Multiple Repetitions}

Multiple repetitions can occur when the stellar system contains $3$
or more masses. 
To derive analytic expressions, 
we first consider lensing by a static \ps\ whose \op\ is aligned with
the \lp . The effects associated with the \pl s' velocities and with
changing the \otn\ of the \op\ are then briefly considered.
The analytic expressions allow one to
make intuitive predictions. In \S 4 we turn to numerical simulations
to derive
detailed results that can be checked against the predictions.

\subsubsection{Multiple Repetitions in the Static, Face-On Approximation}

If a source track crosses through the orbit of a \pl ,
the probability that it will cross within $w_i$ of the
\pl\ itself is approximately $[(2\, w_i)/(\pi\, a_i)]$.
(We assume that the \pl 's motion during the time the
source crosses its orbit can be ignored, and that
curvature effects are also unimportant.)
Define $P_{\bf X}$ to be the probability that the source track will encounter
a specific subset, {\bf X} = $\{\beta_1, \beta_2, \ldots, \beta_k\}$,
of the system's masses, producing an \ev\ with $k-1$ repetitions.
Considering the regime $a>>w_i+ w_j$, the probability of encountering these
objects, and only these objects, is
\begin{equation}
P_{\bf X} \cong {{w_0}\over{w_{0,0}}} \prod_{i \in {\bf X}, i \neq 0}
   \left( \frac{2 w_i}{\pi a_i} \right).
   \label{eq:approx2}
\end{equation}
if the central star is one of the masses encountered.  If the central star is
not encountered, then
a similar expression is derived, but the contribution is suppressed by 
a factor on the order of $\sqrt{q}.$ 
We note that the curvature of the innermost planet in the set
{\bf X} may be significant and can increase the probability
that a track will encounter this planet.
The probability of an \ev\ with $k-1$ repetitions is obtained by
summing over all distinct sets, $X.$ 

The expression above is useful because they clarify the functional form of the
probability of observing a particular combination of $k$ objects:
generally a rough
proportionality to the ratio between the
width of each \pl 's lensing region and its distance 
from the central star. This
makes it simple to estimate the effects of varying the parameters of the
source-lens system.

\tightenlines

\begin{deluxetable}{llllll}
\scriptsize
\tablecaption{Predictions: 
wide-orbit lensing events for known and model systems}
\tablehead{\colhead{System} \tablenotemark{(2)} &
   \colhead{$P_1-P_\odot$} \tablenotemark{(3)} &
   \colhead{$P_2^{sun}$} \tablenotemark{(4)} &
   \colhead{$P_2^{no sun}$} \tablenotemark{(5)} &
   \colhead{$P_3^{sun}$} \tablenotemark{(6)} &
   \colhead{$P_3^{no sun}$} \tablenotemark{(7)} }
\startdata
Known systems: \\
Gl 229       & 19.9  & 0.7  & 0.0     & 0.0     & 0.0 \\
PSR B1620-26 & 10.0  & 0.4  & 0.0     & 0.0     & 0.0 \\
\hline
Model systems: \tablenotemark{(8)} \\
Power-of-2 & 26.0  & 2.2   & 0.037 & 0.017
   & $8.2 \times 10^{-5}$ \\
Power-of-3 & 14.7  & 1.7   & 0.013 & $7.1 \times 10^{-3}$
   & $1.2 \times 10^{-5}$ \\
Power-of-4 & 12.6  & 1.5   & $7.1 \times 10^{-3}$ & $4.5 \times 10^{-3}$
   & $3.7 \times 10^{-6}$ \\
\enddata
\tablenotetext{(1)}{
The computations that produced the results in this table were based on the
analytic approximations discussed in \S 3.
Only encounters with
wide planets were considered; the results were averaged over all inclinations
of the system to the line of sight, and over all angles of approach for the
source track. The
detectability threshold is $A_{min} = 1.34$ (separation of $1.0 R_E$).
No minimum \ev\ duration was required for detectability.}
\tablenotetext{(2)}{55 Cnc and HD 29587 are not listed here because, as can be
seen in Table 1, the orbital separations of the planets in these systems do
not fall into the regime for wide lensing for the values of $D_S$ and $x$ under
consideration.}
\tablenotetext{(3)}{Percentage of isolated events in which a planet is
encountered.}
\tablenotetext{(4)}{Percentage of events in which two encounters
occur, and in which the central star is one of the objects encountered.}
\tablenotetext{(5)}{Percentage of events in which two encounters
occur, and in which neither object is the central star.}
\tablenotetext{(6)}{Percentage of events in which three encounters
occur, and in which the central star is one of the objects encountered.}
\tablenotetext{(7)}{Percentage of events in which three encounters
occur, and in which none of the three objects is the central star.}
\tablenotetext{(8)}{In each of these systems, the planet closest to the sun
is in an orbit of radius $a = 4.8$ AU.  The system is cut off at a 
maximum possible separation of $10^4$ AU.  Thus the power-of-2 system has 12
planets, the power-of-3 system has 7 planets, and the power-of-4 system has 6
planets.}
\end{deluxetable}

\subsubsection{Orbital Inclination}

\label{sec.inc-effects}

In general
the plane of the orbit will be inclined relative to the line of sight.
Let $\alpha$ represent the angle between the normal to the lens plane and the
normal to the orbital plane.
An approach analogous to the one sketched above for the face-on case
can be used.
The polar symmetry of the system is broken, however, since the
projection of an inclined circular orbit on the sky is an ellipse; the
probability of multiple objects acting as lenses depends on the direction from
which the source approaches the lensing system.
For some directions of approach,
the range of impact parameters leading to detectable events is now smaller,
but a larger fraction of events will involve encounters with more than
one object, and so will appear to repeat.
The effect can be intuited by multiplying each factor of
$a_i$ in the above equations by
the geometrical factor 
\begin{equation}
G \propto \left( \
   \frac{1}{\sqrt{\sin^2 \theta + \cos^2 \theta \cos^2 \alpha}}
   \right).
\end{equation}
where $\theta$ is the angle the track of the source makes
with the semimajor axis of
the elliptical projection of any planet's orbit.
Averaging over all possible angles of approach, there 
is a net increase in the
number of repeating events.

As $\alpha$ tends toward $\pi/2$ and we view the \ps\ edge-on, most directions
from which the source can approach can lead only to isolated \ev s. For a small
swath of tracks, however, the probability that the source will pass through the
Einstein radii of several 
\pl s approaches unity. If the \pl s are of equal mass and \ev s can be detected
when the distance of closest approach is $w_i = n\, R_{E,i},$ 
then the probability of an \ev\ in which the \cs\ and all of the \wo\ \pl s
serve as lenses is roughly $n\, R_{E,i}/D,$ where 
$n$ is given by Eq.\ 4, $R_{E,i}$ is the \re\ of a \pl, and
$D$ is the distance between the two most widely separated objects in the \ps .
The probability of \ec s with all but one or all but two of the masses in
the \ps\ are of roughly the same size, typically ${\cal O}(0.001).$ 
Thus, when the \ml\ teams have carefully followed the progress of
thousands of \ev s, they will have found several in which the source track traces the
global structure of a planetary system that gives rise to several detectable 
\ec s. 

The orbital inclination also influences the wait times between \ec s. 
Averaging over all possible
directions of approach should cause a net decrease in the wait times between
encounters for repeating events, and should also produce a dispersion in the
distribution of wait times.

\subsubsection{Velocity Effects}

\label{sec.vel-effects}

The planets that can be detected via \ml\ are generally far enough from
the central star that their orbital speed, $v_{orb},$ is low. If
$a=\mu\, R_E$ then
\begin{equation}
v_{orb} = 7\, {{km}\over{s}}\, \left( \frac{2}{\mu} \right)^{\frac{1}{2}}
   \left[ \left(\frac{M}{M_\odot} \right)
   \left( \frac{10\, \mbox{kpc}}{D_s} \right)
   \left( \frac{1}{x(1-x)} \right) \right]^{\frac{1}{4}}.  \label{planetvel}
\end{equation}
 This shows that the planets that can be discovered
through microlensing, and especially wide planets, tend to be orbiting with
fairly low speed.  Hence, the transverse speed $v_t$ of the source (relative
to the central star of the lensing system) is likely to be large enough
compared to $v_{orb},$ that the static approximation used above can provide a
good guide to the \ev\ probabilities. Nevertheless, there will be a small
number of \ev s in which the relative magnitudes and \otn s of the transverse
source velocity and a \pl 's orbital velocity will influence the
encounter probability or the characteristics of an observed event.
This can occur
when \vt\ is drawn from the low-\vy\ end of the \vy\ distribution,
particulary if $x$ is close to unity (or zero) and the central
star is massive.
For example, if the transverse velocity of the source is
small enough that $R_{E,i}/v_t$ is a significant fraction of the orbital
period of planet $i$, the event probability increases.  This is because the
source spends a nontrivial amount of time within $R_{E,i}$ of the planet's
orbit;  if the planet does not lie in the source's path at the beginning of
this interval, it may move to cross the source's path during the interval,
resulting in a detection.
In addition, \ev\ \du s can be significantly decreased or increased,
depending on the angle between \vt\ and $v_{orb}.$

\section{Simulations of Lensing by a Planetary System}

To include all of the the relevant effects, we have carried out
sets of simulations in which specific planetary systems serve as lenses.
Each simulation was a numerical experiment in which the projections
onto the lens plane of a large number
 of randomly selected source tracks were followed.  
The model planetary systems we have considered are (1) the solar system; (2)
a power-of-3 model with seven Jupiter-mass planets; (3) a power-of-2 model
with twelve Jupiter-mass planets. 
 In all cases we neglected ``inner" planets, i.e.,
those whose physical spatial separation from the central star
was less than $\sim 1.5\, R_E$, where $R_E$ is the
Einstein radius of the star.  All of the orbits have been
taken to be circular, with the motion governed by Newton's laws.  For each
model planetary-lens system, we have placed the source population in the
Galactic Bulge ($D_S$ = 10 kpc),
and the lens at $x\, D_S$ = 9 kpc.  The central star in each of these systems was
chosen to be of solar mass.

At the time when the source started moving along the track with some
velocity, $\vec{v_t}$, we
started the planets in motion, each at a randomly chosen orbital phase.
As time progressed, we tracked the position of the source and of each
of the planets.
We wanted to determine when the source track passed close enough to any mass
in order for there to be a potentially observable ``encounter", roughly
how long each such encounter would last, and how many encounters there would
be as the source traveled along a specific track. Because the details
of the magnification as a function of time are not needed to derive this
information, we used only the value of the projected separation between the
source and each planetary-system mass to determine whether an
encounter was in progress. Specifically, we asked whether the projected
separation was smaller than some pre-selected value,
$w_i=n\, R_{E,i}.$ 
As before, $i$ labels the masses in the \ps, with $i=0$ corresponding to the
star, and $i$ ranging from $1$ to $N$ for the planets.
The values of $n$ to be used were selected at the beginning of each simulation.
[$n=1$ ($n=2$) corresponds to a magnification of $1.34$ ($1.06$).] 

$10^7$ tracks were used to sample the events expected when the solar
system served as a lens, and
$3\times 10^7$ tracks were used to sample the
power-of-$3$ and power-of-$2$ models.  
Because even the larger numbers of tracks used for the 
power-of-$3$ and power-of-$2$ models do not provide 
sampling equivalent to that used for the solar system, we have
normalized the results so that the effective linear density of
source tracks was the same for each \ps .  We have also smoothed the
power-of-$2$ and power-of-$3$ distributions to remove effects  
due to poor sampling. 
We note, however, that the finer 
features of the statistical distributions, i.e., those which become
apparent only as the number of \ev s  increases, are best seen in the plots 
of the solar system distributions. 
 
We have used our simulations to derive in detail what happens for
each planetary system, and also to test the effects of varying
some of the event parameters and detectability criteria.

\subsection{Event Parameters}

The characteristics of the \lc s
are determined by the characteristics of the event.
In particular, we must consider (i) the orientation angle, $\alpha,$
between the plane of the orbit and the lens plane, and (ii) the transverse
speed, $v_t,$ of the source with respect to the central star of the planetary-system
lens.
We have therefore carried out some simulations in which we have varied $\alpha$ and $v_t,$ in
order to test their influence on the results.
In more realistic simulations, $\alpha$ was chosen uniformly over the
interval $0-2\, \pi$, and $v_t$ was chosen from a Gaussian distribution
centered at $150$ km/s, with width equal to $50$ km/s.

\subsection{Detectability Criteria}
\label{detectcrit}

As presently implemented, the search for resonant planets is carried out in
two steps.  First, the monitoring teams (e.g., EROS, MACHO, OGLE) 
identify \ml\ \ev\
candidates.  They monitor the flux from {\cal O}($10^7$) stars regularly.
Some of the monitored fields are not visited every night; others
are the
targets of regular nightly monitoring and may occasionally be re-visited
even within a single night. 
If the monitoring teams discover that an otherwise non-variable star's flux
has increased significantly above baseline in several consecutive
measurements, they typically issue an alert so that other observers can
monitor the star more frequently. Follow-up teams have been formed to take
systematic advantage of this opportunity
(Udalski et al. 1994; Pratt et al. 1995; Albrow et al. 1996).
Under
favorable conditions, the follow-up teams carry out hourly
monitoring with good photometry ($\sim 1\%$).

Within this framework, there 
are two key elements of detectability.
First, does an encounter last long enough to be
detected?
Even in principle,  an encounter cannot be detected
unless it 
is caught in progress during at least one observation.  
Reliable detection generally requires the \ev\  to   
last long enough to span at least the time interval
between two or more consecutive observations.
In some of our simulations, we have assumed that a minimum \ev\ duration of
one day is needed in order to reliably detect the first encounter between a
source track and the Einstein ring of a lens.  This criterion 
can be achieved by the \mo\ teams in some fields,
using their present observing
strategy. 
In others simulations,
 we have
dropped the requirement of a minimum duration for the first event.
This more relaxed condition is appropriate to the follow-up teams;
since they can achieve hourly \mo , even an \ev\ lasting $8$ hours
(such as one likely to be due to an Earth-mass \pl), 
can be
readily identified if the follow-up teams  attempt to discover new \ev s.

The second key element of detectability is provided by the value of the
peak magnification: 
what is the minimum
peak magnification, $A_{min},$ required in order to reliably determine that
an \ev\ occurred?
When the observing teams started, they tentatively chose $A_{min}=1.34,$
corresponding to a distance of closest approach equal to $R_E$.
It turned out, however, that some apparent events with magnification above
$1.34$ but less than $\sim 1.58$ were due to stellar variability; this has
led the MACHO team, for example, to use $A_{min}=1.58$. It is likely, however,
that this condition can be relaxed as the continued study of the
same fields over time will allow for better identification and tracking of stellar
variability and decrease its possible contamination of our count of
true \ml\ events. In principle, the value of $A_{min}$ is set by the
photometric precision of the monitoring system.
If the photometry is good to the $1-2\%$ level, 
then smaller values of $A_{min}$ are achievable. 
 $A_{min} = 1.06$ corresponds
to a distance of closest approach 
approximately equal to $2.0 R_E$, and  $A_{min} = 1.02$
corresponds
to a distance of closest approach 
approximately equal to $3.0 R_E$. (To achieve this latter
value would require better photometry than is typical of even the
present-day follow-up teams.) 
We have carried out two types of simulations. In the first, we have assumed
that \ev\ identification was being done by the \mo\ teams; in these, we
assumed that the distance of closest approach needed to be at least as
small as $R_E$. In the second, we have assumed
that \ev\ identification was being done by the follow-up teams; in these, we
assumed that the distance of closest approach needed to be at least as
small as $2\, R_E$. 
We note that finite source size effects can make \ev s
detectable when the distance of closest approach is even larger. We
return to this point in \S 7; the results presented in this section
  were derived
under the assumption that the size of the lensed source could be
neglected.

We label
the three sets of detectability criteria used in our simulations:
``A'', ``B'', and ``C''.  

\noindent Criteria A:  The first \ec\ must exhibit magnification greater than  
	$A = 1.34$ (source-lens separation less than $1.0 R_E$) for at least
	1 day in order for the lensing \ev\ to be detected.  
        After the detection of a first \ec , 
	subsequent \ec s can be
	detected when $A > 1.06$ (source-lens separation less than $2.0 R_E$),
	and are not subject to minimum duration requirements. 

\noindent Criteria B:  All lensing \ec s are detected when the 
        source-lens separation becomes smaller than 
	$2.0 R_E$. The first encounter must have a duration of at least 1 day
         in order to be detected,
	but after one \ec\ has been detected,
        there is no minimum duration required for the \dtn\
        of subsequent \ec s.

\noindent Criteria C:  All lenses are detected at a source-lens separation of
	$2.0 R_E$, and no minimum duration is required in order
        to detect any encounter.

Because we want to make contact with observations, we have defined \ec\ \du s
and wait times to reflect the actual required monitoring times
during and between \ev s. 
The duration of an encounter is always defined to be the interval of time
during which the source was within $2\, R_{E,i}$ of a lens.
Wait times are defined as the time intervals between encounters. 

We have used the normalization of \S 3.2.
The primary difference is that we express the rates in terms of 
percentages.
$P_1$ is the percentage of source tracks that cross through the \er\ of a
single object. 
All percentages are computed with respect to the number of isolated
stellar lens \ev s
(with $A_{peak}>1.34$) in our 
simulations. 
\footnote{Note that the exact normalization described in \S 3.2
requires dividing by the number of \ec s, rather than by the number of
isolated \ev s in which the \cs\ serves as a lens. Because these two different
ways of computing the rates lead to similar results (using the number of
\ev s yields rates that are $\sim 2-10\%$ higher), and because
the number of \ev s provides a more straightforward comparison,
we have chosen to divide by the number of isolated stellar-lens \ev s.} 
We also frequently refer to these percentages as ``rates", since,
as discussed in \S 3.2, they should be roughly equal to the number of
\ev s of the given type expected per year along the direction to the Bulge.  

$(P_1-P_\odot)$ is the percentage of source tracks passing through
just one object, excluding cases in which the \cs\ is \ec ed. $P_{1,\ov}$
is the percentage of source tracks which 
in which the influence of \ec s with $2$ lenses
is clearly
visible in the \lc, but the \lc\ exhibits 
just one continuous perturbation 
(i.e., once the \mage\ falls below 
the detectability limit, it does not again
rise above it).
$P_i$ is the percentage of source tracks crossing through
the Einstein radii of $i$ objects. $P_{i,\ov}$
is the percentage of source tracks passing through $i+1$ Einstein
rings, but in which
the presence of two of the lenses (the \cs\ and the innermost \pl\ in a \wo ) 
is
detected in an \ov\ \ec ; the \mage\ rises above and falls below
the detectability limit  
only $i$ times.

\subsection{Results}

\subsubsection{The Effects of Systematic Variation of $v_t$} 

\begin{figure}
\vspace{-1 true in}
\plotone{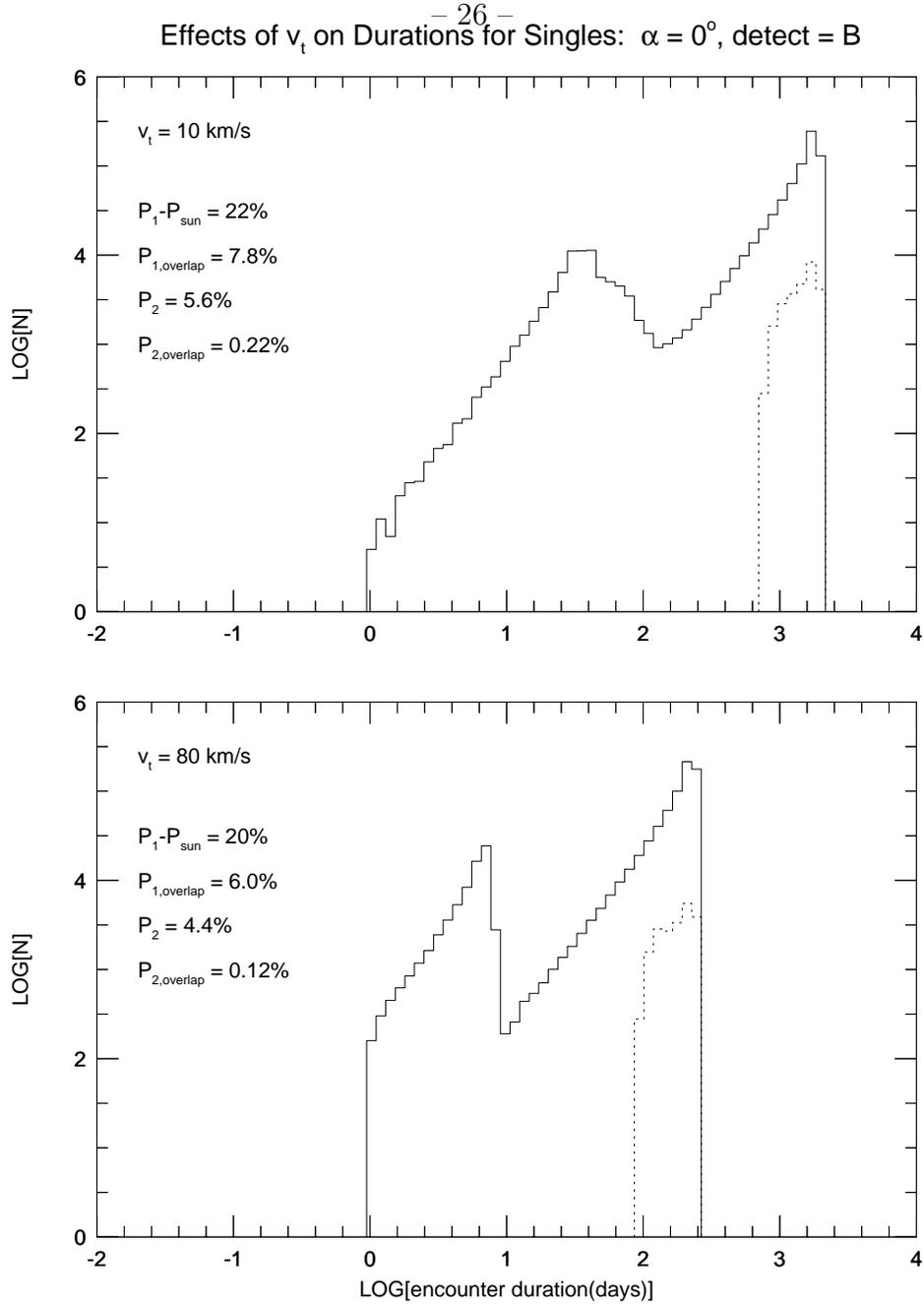}
\vspace{-0.5 true in}
\caption{Distribution of encounter durations for isolated \ev s. 
The duration is defined to be the time interval during which the projected
separation between the source track and the lens is smaller than $2\, R_{E,i}$,
where $R_{E,i}$ is the Einstein radius of the lens.  
The \ps\ is a power-of-$2$ 
system with five Jupiter-mass planets; the orbital plane was coincident 
with the lens plane ($\alpha=0$).  Top panel: $v_t = 10$ km/s.
Bottom panel $v_t = 80$ km/s.  
The dashed curves show the duration of overlap \ev s. 
}
\end{figure}

We used a truncated (5-\pl) power-of-2 system to explore the influence
of changing $v_t$ on the rate of events and characteristics of \ev s.
\footnote{The reason for the truncation was simply to achieve 
better sampling, i.e., a higher density of source tracks.} 
In the simulations whose results are shown in Figure 4, we took 
$\alpha=0;$ i.e., the orbital plane coincided with the lens plane.    
Holding $\alpha$ fixed, we systematically varied $v_t.$ 
The distribution of \ec\ \du s for $v_t=10$ and $80$ km/s are shown in Figure 4.
A comparison between the two cases clearly illustrates that, in keeping 
with the predictions,
the 
overall rate of detected \ev s is larger for all types of \ev s when
$v_t$ is smaller. This effect is most pronounced for repeating events
which involve several planets, since the probability of a repeating event
behaves like a product of detection probabilities for each planet separately.
The percentage $P_{2,overlap}$ of overlap doubles demonstrates this well,
since an overlap double involves at least three objects, two of which must be
planets.  $P_{2,overlap}$ is larger for $v_t = 10$ km/s
by a factor of $\sim 1.8$. 
When $v_t$ is comparable to the orbital velocity, as it is in the
top panel, there is a clear dispersion in the distribution 
of encounter durations corresponding to short-duration \ec s in which
a \pl\ serves as a lens.  This dispersion is due to the influence of the 
\pl s' orbital velocities; encounters can be lengthened or shortened depending
on the angle between the planet's orbital motion and the transverse source
velocity.
In contrast, for $v_t = 80$\  km/s,
the peak due to planets is sharp and well-defined.  

As $\alpha$ increases, the projection of the orbits onto the lens plane
become ellipses. The projected orbital speed along the
semi-major axis is the same as before, but the projected orbital speed 
along the transverse direction is smaller. Thus, along some directions of
approach the effects associated with the finite size of $v_t$ become 
less pronounced.

\subsubsection{Systematic variation of orbital inclination}

\begin{figure} 
\vspace{-1 true in}
\plotone{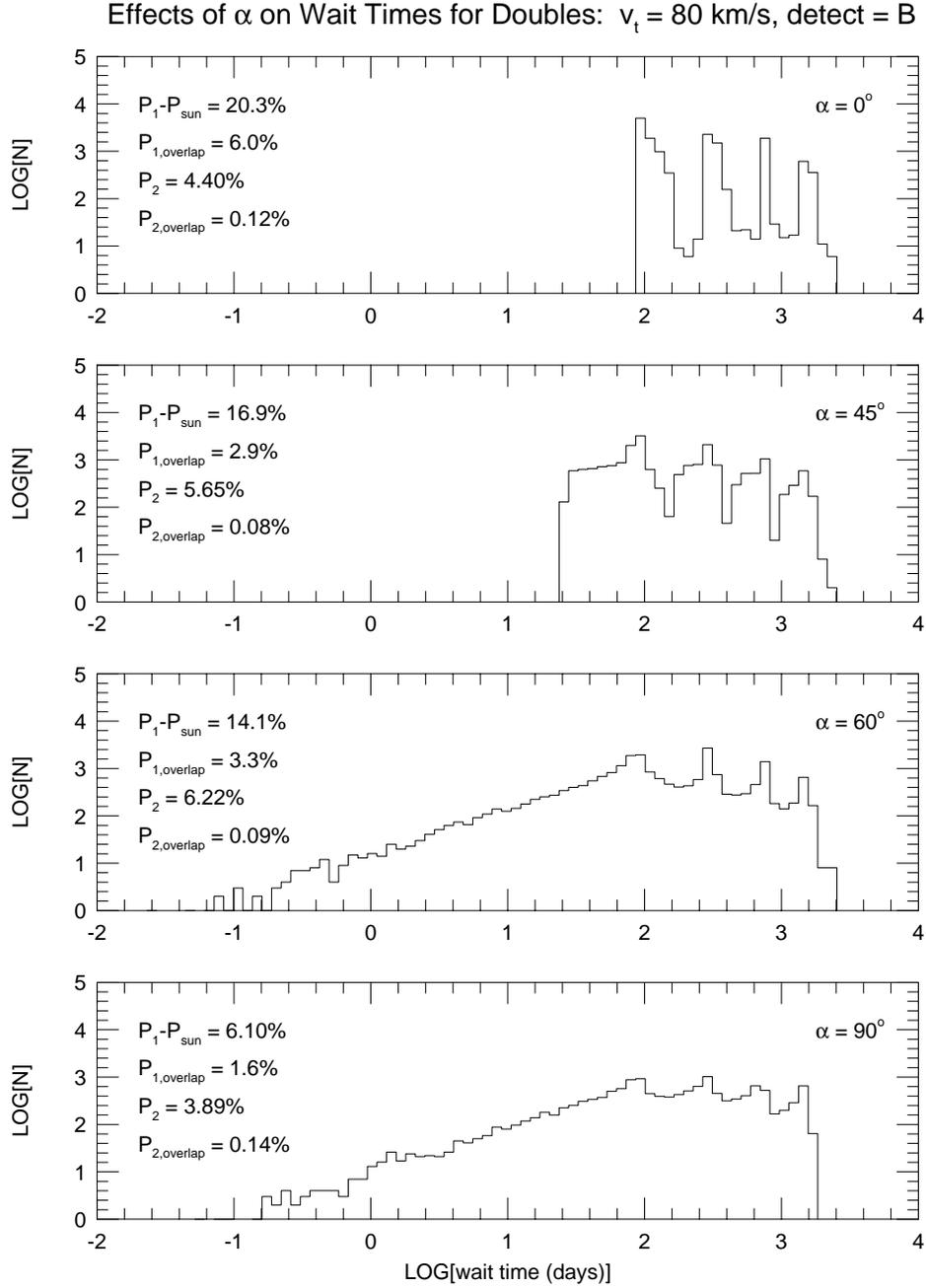}
\vspace{-0.5 true in}
\caption
{Distributions of durations of and
wait-times between encounters for events with one
repetition.  The model \ps\ serving as a lens was the 5-planet power-of-$2$
system.}
  
\end{figure}

To systematically
 test the results of changing $\alpha,$ we again used the 5-planet
power-of-$2$  system, this time keeping $v_t$ fixed at $80$ km/s.
The results for $4$ 
simulations ($\alpha=0^\circ, 45^\circ, 75^\circ,$ and $90^\circ$)
are shown in Figure 5.
For $\alpha = 0^\circ$, four peaks in the distribution of wait times
between \ec s are sharp and
clearly visible; these correspond to wait times between encounters with the
central star and encounters with one of the four outermost planets.  (The
innermost planet is so close that it serves as a lens only in overlap events.)  The
even spacing between the peaks is a signature of the power-of-2 model,
becoming about $0.3 \approx \log_{10} 2$ for planets with $a_i >> R_{E}$.
As the inclination increases, the dispersion mentioned in \S 3.         
appears; note that increasing inclination can only decrease wait
times, so that each peak is ``smeared'' out to the left until at high
inclinations a wedge-like shape is achieved.
The detection rates for non-repeating
events decreased with increasing inclination, but the relative 
detection rates for
repeating events increased, as predicted.
For $\alpha = 45^\circ$, the innermost planet's orbit comes within
$0.7 R_{E}$ of the star, placing it in the \zres\ most of the time.
The detection rates for doubles and overlap singles decrease
dramatically because of this, but they rise again at $\alpha = 60^\circ$ as
the second planet's projected orbit becomes small enough for overlap
\ev s to occur.  
At $\alpha = 90^\circ$ (edge-on), the motion of the \pl s 
brings all of them into the \zres, or even closer, part of the time.
This 
decreases overall detection rates for \pl s
in \wo s. The detection rates for
doubles decrease dramatically; this is because all planets lie along the same
line, and events which might have been doubles actually become triples,
quadruples, or higher-order events.
In fact, we found that for   
$\alpha = 90^\circ$, the percentages $P_3$, $P_4$, and $P_5$ were all approximately
0.3\%. 
Thus, as mentioned in \S 3, 
a relatively large fraction of \ev s can exhibit multiple repetitions,  
as the track of the source sweeps across the ecliptic, crossing through
the Einstein ring of several \pl s and the central star.

\subsubsection{The Effects of Changing the Detectability Criteria}
\tightenlines

\begin{deluxetable}{lllllll}
\scriptsize
\tablecaption{Simulation Results:
 wide-orbit events for known and model systems}
\tablehead{\colhead{Detect \tablenotemark{(1)} } &
   \colhead{$P_1-P_\odot$ \tablenotemark{(2)} } &
   \colhead{$P_1^{overlap}$ \tablenotemark{(3)} } &
   \colhead{$P_2$ \tablenotemark{(4)} } &
   \colhead{$P_2^{overlap}$ \tablenotemark{(5)} } &
   \colhead{$P_3$ \tablenotemark{(6)} } &
   \colhead{$P_3^{overlap}$ \tablenotemark{(7)} } }
\startdata
\multicolumn{7}{l}{\bf Solar system, $V = 150$ km/s,
   $\alpha = 0^\circ$:} \\ \\
A & 0.3 & 2.2 & 0.5  & 0.01 & 0.0      & 0.0 \\
B & 1.8 & 6.0 & 1.4  & 0.04 & $3.0 \times 10^{-4}$   & 0.0 \\
C & 4.4 & 6.0 & 1.7  & 0.05 & $2.4 \times 10^{-3}$   & 0.0 \\
\hline
\multicolumn{7}{l}{\bf Solar system, $V = 150$ km/s,
   $\alpha = 75^\circ$:} \\ \\
A & 0.1 & 1.7 & 0.8 & 0.02  & $1.8 \times 10^{-3}$
   & $1.5 \times 10^{-4}$ \\    
B & 0.3 & 2.4 & 1.7  & 0.04  & $4.2 \times 10^{-3}$
   & $1.5 \times 10^{-4}$ \\
C & 1.5 & 2.4 & 2.3  & 0.04  & 0.01
   & $1.5 \times 10^{-4}$ \\
\hline
\multicolumn{7}{l}{\bf Solar system, $V =$ Gaussian,
   $\alpha = $ uniform:} \\ \\
A & 0.3 & 2.1 & 0.7  & 0.02  & $1.4 \times 10^{-3}$ & 0.0 \\
B & 1.4 & 4.2 & 1.8  & 0.04  & $3.2 \times 10^{-3}$ & 0.0 \\
C & 3.1 & 4.2 & 2.1  & 0.05  & $5.5 \times 10^{-3}$
   & $1.5 \times 10^{-4}$ \\
\hline
\multicolumn{7}{l}{\bf Power-of-2 system, $V =$ Gaussian,
   $\alpha = $ uniform:} \\ \\
A & 23.7 & 2.2 & 3.0  & 0.03  & 0.03   & 0.0 \\
B & 53.8 & 4.0 & 6.6  & 0.08  & 0.06   & 0.0 \\
C & 55.9 & 4.0 & 6.6  & 0.08  & 0.08   & 0.0 \\
\hline
\multicolumn{7}{l}{\bf Power-of-3 system, $V =$ Gaussian,
   $\alpha = $ uniform:} \\ \\
A & 13.0 & 1.9 & 1.5  & 0.03  & 0.01   & 0.0 \\
B & 29.3 & 3.6 & 3.1  & 0.06  & 0.02   & 0.0 \\
C & 30.4 & 3.6 & 3.1  & 0.06  & 0.02   & 0.0 \\ 
\enddata
\tablenotetext{(1)}{Descriptions of the detection criteria can be found in
the text.  All probabilities are given as percentages of the number of events
in which the central star 
was the only lens encountered, and in which the magnification
reached at least $A_{min} = 1.34$.  
}
\tablenotetext{(2)}{Percentage of isolated (non-repeating) 
events (one ``peak'' in
the light curve). Only a single \pl-lens was \ec ed.} 
\tablenotetext{(3)}{Percentage of non-repeating events 
which exhibited
evidence of lensing by two masses. These are ``overlap'' \ev s;
in these cases the two lenses were almost 
always the central star and the innermost
\pl .}
\tablenotetext{(4)}{Percentage of events with one repetition
(two ``peaks'') These are not overlap \ev s; two well-separated masses were
\ec ed. In most cases these two masses are the central star and
the second \pl\ out, but there are other contributions as well.}
\tablenotetext{(5)}{Percentage of single-repetition 
events in which one component
consisted of overlapping \ec s; 
the repetition was due to lensing by a third mass.}
\tablenotetext{(6)}{Percentage of events with two repetitions
(three ``peaks''); all lenses were well separated, with no overlap.}
\tablenotetext{(7)}{Percentage of double-repetition events
in which one component
showed evidence of 
overlapping \ec s; the two ``repetitions" were due to lensing by two other
well-separated masses. If the linear density of source tracks passing through
the power-of-2 and power-of-3 systems had been the same as for the 
solar system, there would have been \ev s in these categories.}
\end{deluxetable}

\begin{figure}
\vspace{-1 true in}
\plotone{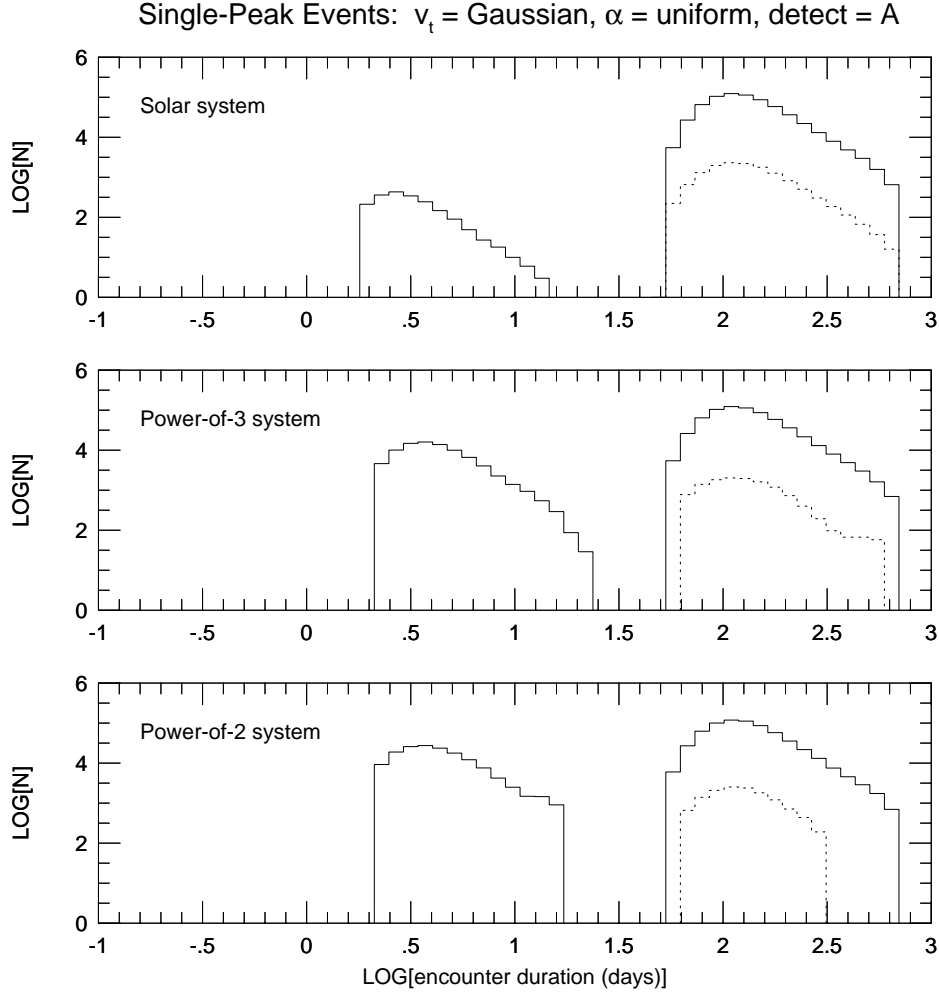}
\vspace{-2 true in}
\caption{Distributions of durations for isolated (non-repeating) events
using the detection criteria of set A.  
In each panel, the peak on the left is due
to encounters with planets (for the solar system, mainly Saturn with a few isolated events from Jupiter), and the peak on the right is due to
the central star.  Overlap events, produced almost exclusively by 
\ec s involving the innermost wide planet and the central star,
are shown by the dotted lines.
The truncation of the large-duration end of the overlap distributions 
in the power-of-3 and power-of-2 models is due solely
to the poorer statistics achieved in the simulations of these models; 
all of the distributions were normalized to the linear sampling achieved
for the solar system.  
}
\end{figure}

\begin{figure}
\vspace{-1 true in}
\plotone{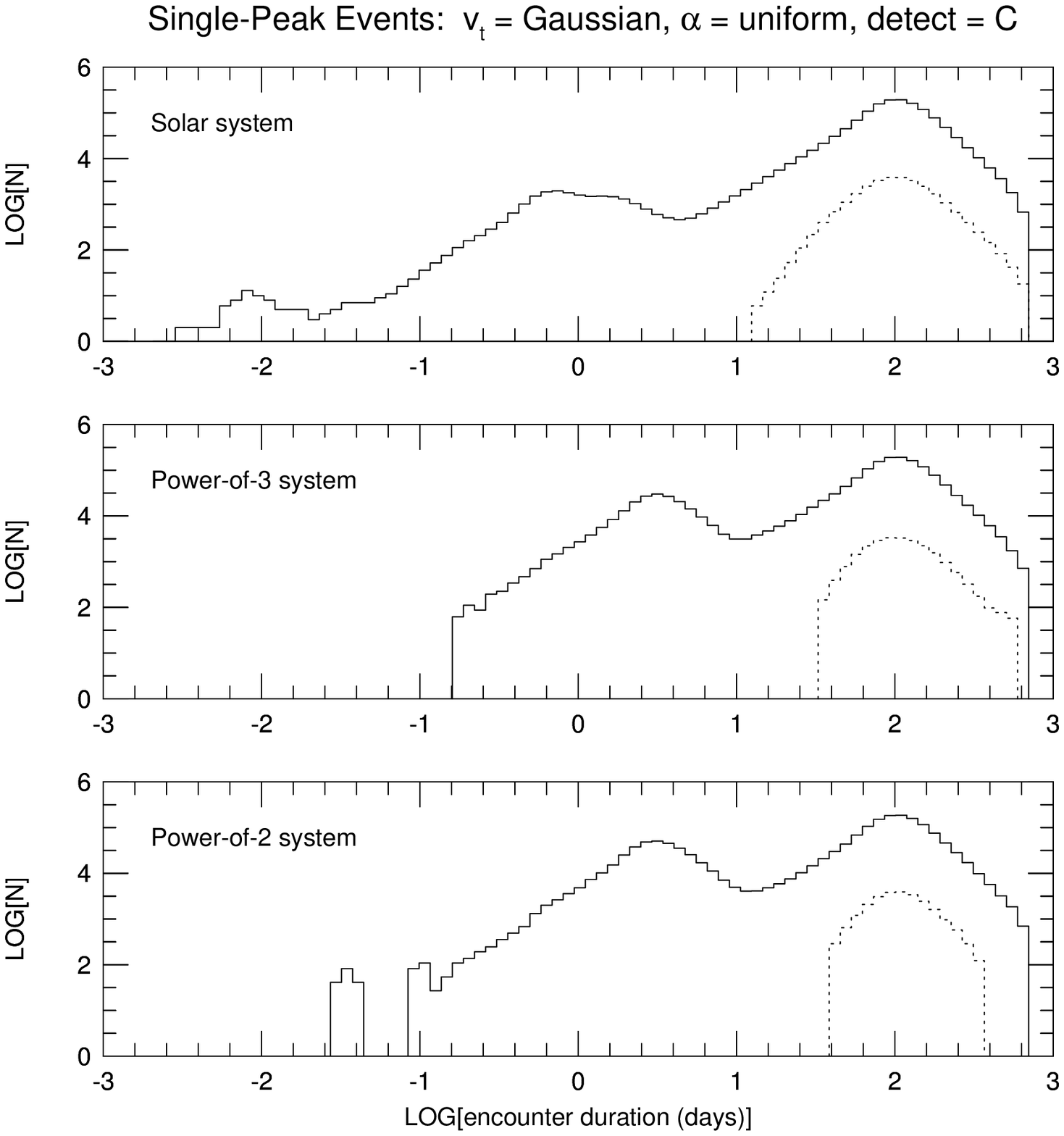}
\vspace{-2 true in}
\caption{Distributions of durations for isolated (non-repeating) events
using the detection criteria of set C.  Here,
because an \ec\ did not have to have a minimum \du\ in order to be detected, 
 the peaks tend to 
have low-duration
tails which can run together. 
The right-most peak in each panel is due to the central star of the system,
with overlap events involving the innermost wide planet being shown by
dotted lines; the peaks on the left are due to planetary encounters.
The central feature in the top panel is a superposition of peaks corresponding
to Uranus, Neptune, and Saturn; the encounters due to Saturn are visible as a
shoulder at about 1.6 days.}
\end{figure}

To better learn how to 
optimize the returns from the \ml\ observations, we tested the effects
of varying the detectability criteria, from the most conservative set
of criteria (set A), to the most inclusive (set C). The results
are shown in Table 3, and in Figures 6--10. 
 
The planetary systems serving as lenses are  
our solar system, the twelve-planet
power-of-2 system, and the $7$-\pl\ power-of-3 system.  
Figures 6 and 7 show the
duration distributions for non-repeating events obtained 
when using detection criteria A and C, respectively. 
Figures 8 and 10 show the
duration and wait time distributions for events with one repetition
when using detection criteria A and C, respectively.
Figure 9 shows the cumulative distributions for the graphs plotted in
Figure 8.

The detection 
criteria of set A are the most restrictive.  When our solar system
serves as a lens, with
$v_t$ = 150 km/s, only Jupiter and sometimes Saturn are able to produce events
sustaining a magnification of $A_{min} = 1.34$ for longer than 1 day.  Any
other planets would only be detected by the follow-up teams, i.e., only if a
larger object (almost always the Sun) was encountered first.  Thus, 
even though a realistic velocity distribution will help, wide-orbit 
planets which are slightly less massive than Saturn are difficult to detect
using a strategy based on a set of criteria similar to that of set A.  
The power-of-2 and power-of-3
models we examined contained only Jupiter-mass planets, so each planet was
detectable,
except for the small fraction of \ec s in which the source track just
grazes the Einstein ring of the \pl . 
 We note, however,
that the criteria of set A are not powerful tools for the
discovery of solar systems  like
our own.

Using the detection criteria of set B (detecting even a first \ec\ 
at $A = 1.06,$ instead of 
$A = 1.34$)
generally more than doubled the detection frequencies in our most
realistic simulations, in which $v_t$ was chosen from a
Gaussian distribution and $\alpha$
was chosen from a uniform distribution.  
Planets could be detected at a lower peak magnification, and more
planets in the system were able to sustain a magnification of $A = 1.06$ for
the 1-day minimum duration required of the first encounter.  For the solar
system, Saturn was often detected as an isolated lens, or as the
first lens \ec ed, and Uranus and Neptune made occasional
appearances as isolated lenses.  
Note that an increase by a factor of $2$
is expected, since, even with the set of detection criteria A, $n$ was
already equal to $2$ for the second \ec . 
Deviations from the factor of $2$ increase
 are mostly associated with \ev s which include
\ec s with a low-mass \pl, and are therefore seen primarily in the 
solar system; these deviations are due to our use of a minimum
duration for the  detection of the first \ec .

Finally, the detection
criteria of set C (removing the requirement that the first encounter last
for at least a day) made Uranus and Neptune more regularly
 detectable as isolated lenses,
whereas before they were primarily detected as part of repeating events 
involving a larger mass.  These planets produce significant new structure in
the duration distribution of Figure 7.  Even Pluto produces its own peak at
$10^{-2}$ days ($\sim 10$ minutes), although this is too short to be
detected in practice.  

To assess the relative benefits of the three detection strategies used,
we note that the largest increase in the rates of detectable planet-lens 
\ev s was realized by switching from the set of criteria A, in which $A_{min}$
for the first \ec\ is $1.34$ to set B, in which $A_{min}$
for both \ec s is $1.06.$    
The gains in switching from B to C, which eliminated the 
requirement that the first detectable \ec\ have a \du\ of 1-day,
could be substantial only if, like our own solar system, the lens
\ps\ contains wide-orbit \pl s less massive than
Jupiter.

In general, the smaller the value of 
$A_{min},$ the longer the duration of the observed \ev\ anyway.
Thus, while frequent \mo\ is desirable, and more frequent \mo\ than is
presently achieved by the \mo\ teams would be very valuable, 
the key issue is sensitivity to \ec s in which the peak \mage\ may
be smaller than $1.34.$ We note that this result is likely to remain
valid, even when other effects are considered. For example, 
both blending and \fsse\ decrease the peak \mage . 
Thus, detection rates for \ev s subject to these effects 
may also be improved if \ec s which achieve
only smaller values of $A_{min}$ can be detected.

\begin{figure}
\vspace{-1 true in}
\plotone{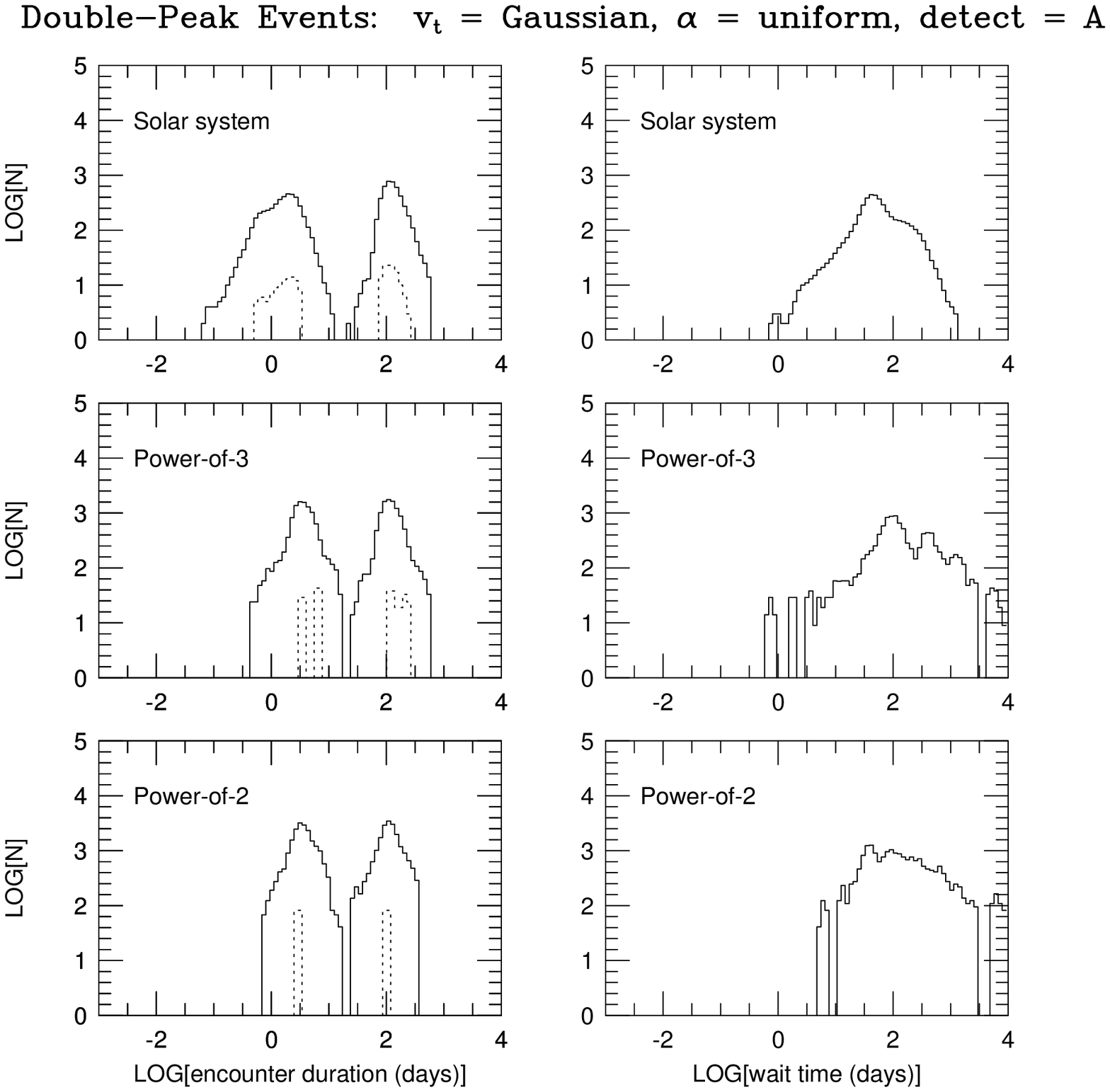}
\vspace{-2 true in}
\caption{Distributions of \ec\ durations and of and wait times 
between \ec s for events with one
repetition, using the detection criteria of set A.  
{\bf Durations:}\  For the solar system (top left),
Saturn dominates the left-most peak, with Uranus and Neptune
appearing as a shoulder
to the left, at approximately 0.7 days.  Overlap doubles are shown with a
dotted line:  overlap \ec s involving both Jupiter and the Sun 
are represented in the
right-most peak; the other \ec\ that occurred as part of these
repeating \ev s was with either Saturn, Uranus, or Neptune, and  
produced the peak on the left.
{\bf Wait times:}\ For the solar system, the predominant peak
is produced by doubles involving the Sun and Saturn; Uranus and Neptune
produce a broad shoulder to the right, at around $10^2$ days.}
\end{figure}

\begin{figure}
\vspace{-1 true in}
\plotone{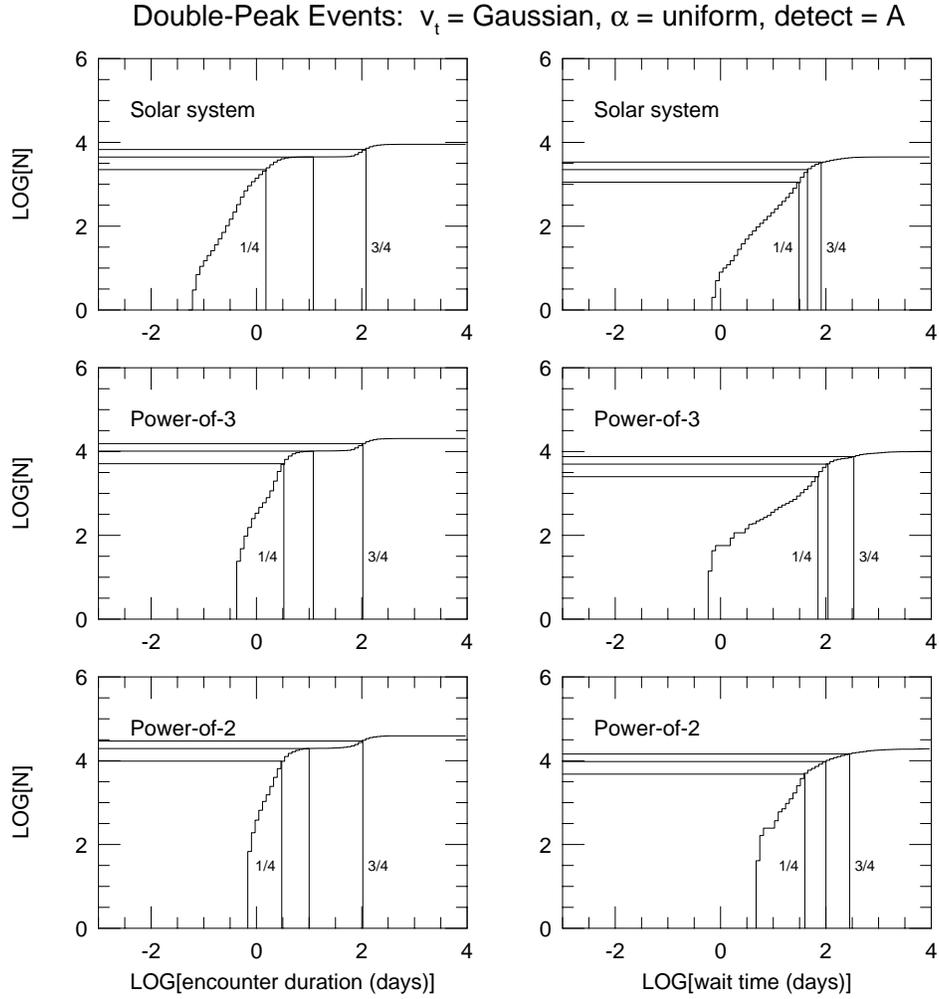}
\vspace{-2 true in}
\caption{Each panel shows the integrated area under the distribution to which
it corresponds in the previous figure.  
Encounter durations are shown on the left; wait times between
encounters are shown on the right.  Roughly $75\%$ 
of all encounters, and half of
the gaps between encounters ($75\%$ for the solar system), last less than 100
days.}
\end{figure}

\begin{figure}
\vspace{-1 true in}
\plotone{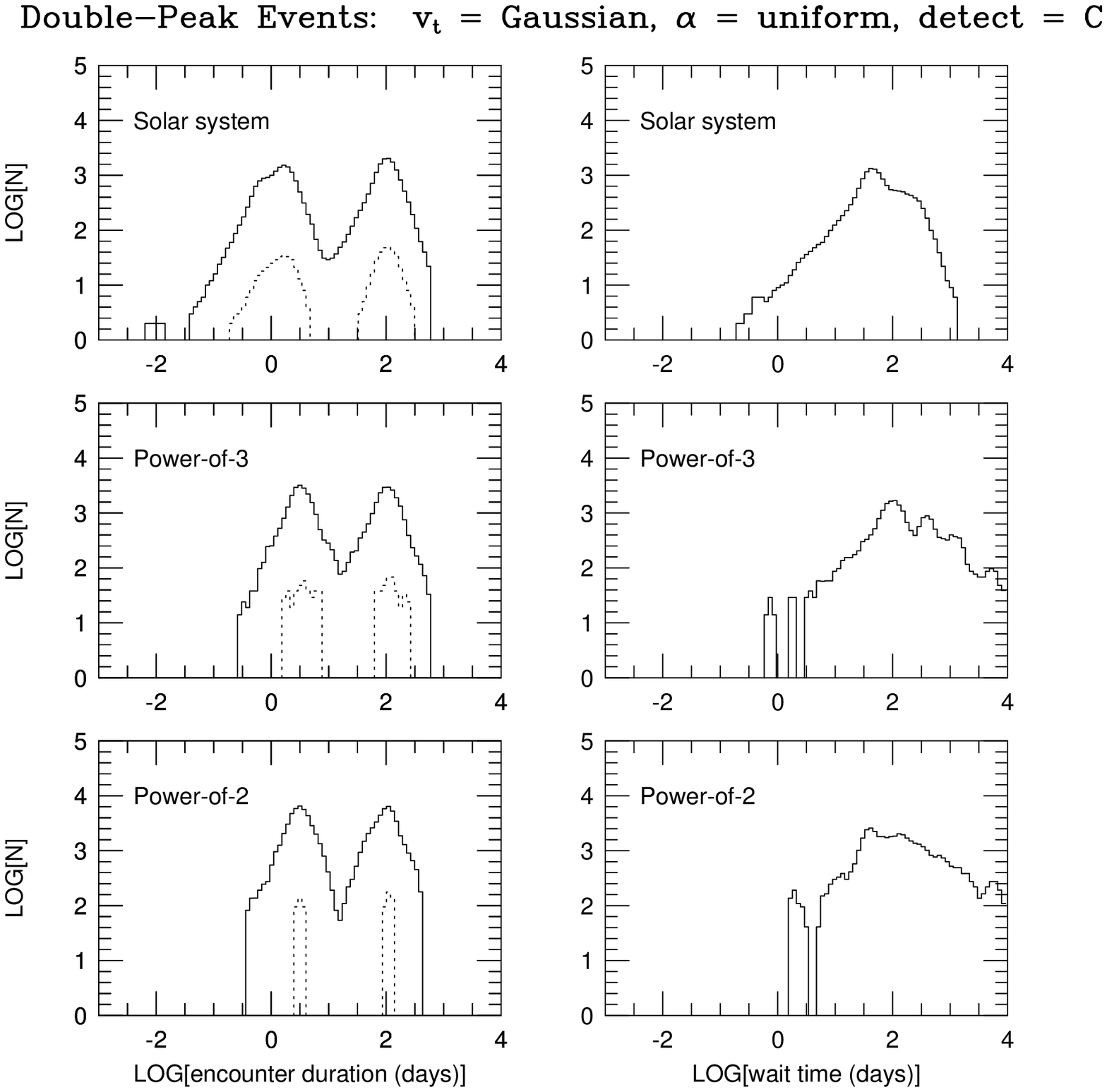}
\vspace{-2 true in}
\caption{Distributions of durations of and wait times between events with one
repetition, using the detection criteria of set C.  
Most of the structure visible in Figure 8, where
the detection criteria of set A were used, can also be seen here.
The overall event rates have
increased, and the gaps in the centers of the distributions have filled in.
A new peak corresponding to Pluto has appeared in the duration distribution
for the solar system (top left); these \ev s are so short that, although we
may occasionally catch one magnified point on their \lc s,
 we are unlikely
to be able to follow them. }
\end{figure}

\subsubsection{General Features}

Perhaps the most interesting feature of the results is that 
they clearly indicate the feasibility of the search for \ps s 
containing \pl s in \wo s. 
For the most optimistic model, the 12-\pl\ power-of-$2$ model,
even the strictest detection criteria
yielded a rate 
 of  $4.5\%$
 for \ev s showing some
evidence of two or more objects in the system and, $23.5\%$  
for isolated short-duration \ev s. 
\footnote{These rates are computed using the normalization
described in \S 3.2 and \S 4.2. That is, they are $100$ times the number
of \ev s of the type described, divided by the number of \ev s in which
the central star served as the lens in an isolated \ev, with the
peak \mage\ of the stellar \ev\ achieving a value greater than $1.34.$} 
For the detection criteria of set C, $10\%$ was the rate of \ev s showing some
evidence of two or more objects in the system, and $\sim 56\%$
was the rate of isolated short-duration \ev s.  
It is clear that, if the power-of-2 model is 
realized in nature with any great
frequency, we will be able to observe many \pl-lens \ev s even
during the next few years. Conversely, an absence of large numbers of
interesting \pl-lens \ev s, particularly if we use the detection criteria
of set C, would allow us to definitively falsify the hypothesis that most
stars are accompanied by power-of-$2$ \ps s. 

While 
it may not be surprising that such a radical model is verifiable
or falsifiable, the relative ease with which \ps s such as our own
can be discovered is indeed worthy of note. The rates of \wo\ \ev s   
range from $\sim 3\%$ for the detection
criteria of set A to $\sim 10\%$ for the detection
 criteria of set C.
This means that, even over the short-term, observing strategies
such as the ones we propose in \S 9, will lead to significant results.
In \S 10 we discuss the likely results for \ps s not modeled in this section. 

An important question for the observing teams is how long they must
continue frequent \mo\ in order to have a good chance of detecting
a repetition. Figure 9 indicates that 
$1/2-3/4$ of all \ev s with one repetition
will begin the repetition within $100$ days of the time the 
magnification associated with the first \ec\ dips below $1.06$.
In addition, 
a large fraction ($\sim 75\%$) of  overlap \ev s may be identifiable.  
In our simulations, overlap \ev s form a significant
subset ($1-4\%$) of all \ev s for all of the models (solar system
through to power-of-$2$) and for all detection criteria (sets A through C).
This may be an artifact however, because in each case the model
system happened to have a \pl\ located just outside the zone for
resonant lensing. Note though, that if overlap \ev s are not seen
at roughly this level, there are likely to be few resonant \ev s,
since the \pl s that serve as lenses in overlap \ev s are the 
ones that can be brought into the zone for resonant lensing 
as the orientation  of the \ps\ changes. 
    
\section{Microlensing and the Search for Extraterrestrial Life}

Recent and ongoing advances in technology have led to the discovery of
extrasolar planets
(see references listed in 
the Encyclopedia of Extrasolar Planets,
www.obspm.fr/darc/planets/encycl.html),  
 and promise the discovery and even the imaging of
additional planets (Angel \& Woolf 1997; Fraclas \& Shelton 1997; Labeyrie
1996; Brown 1996).  These developments excite the imagination because
they seem to bring us closer to the possible discovery of extraterrestrial
life.  It is therefore interesting to ask whether microlensing searches are
likely to find planets on which life could thrive.

We do not yet have a clear enough understanding of the nature of life to
definitively answer this question, because the range of physical conditions 
compatible with life may well be wider than our limited experience would
at first suggest.  It has been proposed, for example,
that life
may exist on the outer planets of our own solar
system and/or on their moons. (See, e.g., 
Reynolds et al. 1983; 
Raulin {\it et al.} 1992; 
Sagan, Thompson,  \& Khare 1992;  
Williams, Kasting, \& Wade 1997, McCord {\it et al.} 1997.)  
It has even been postulated that life may exist in non-planetary
environments, including the interiors of stars and molecular clouds (see,
e.g., Feinberg \& Shapiro 1980).
Thus, any \pl , however close to or far from its star, and whatever
the nature of the star, can be considered as a possible harbor for life.

Nevertheless,
in the absence of real information on the existence of life away from our
own planet, one question is clearly interesting:  will microlensing find
evidence of planets similar to Earth?  We must of course define what we mean
by ``similar to Earth".  If we mean that there is a chance that chemical
processes necessary for Earth-like life could occur, then we want to
consider planets which can have similar surface and atmospheric make-up, and
similar amounts of energy available to fuel the necessary chemical processes.
Thus we may ask:  by
how much can we change the mass of the Earth, $M_\oplus$, and
its distance from the sun,
$a_\oplus$ (i.e., the quantities microlensing events can measure) and still
maintain conditions that seem able to support life forms with chemistry
similar to that of Earth life?

The range of planetary masses and distances from a solar-type star 
compatible with Earth-like conditions, particularly the presence of
liquid water, has been dubbed the ``Goldilocks Problem", and has
been studied by many researchers. (See, e.g., Rampino \& Caldiera 1994,
for an overview.) Here we make some general observations.
 
The primary requirement on $a$ is that the incident flux of radiation from the star
should be comparable to the flux received by the Earth from the Sun.
That is, ${\cal F}/{\cal F}_\oplus$, should not be too different from unity.
\begin{equation}
{{\cal F}\over{\cal F}_\oplus}={{0.012}\over{\mu^2}}
\left( {{L}\over{L_\odot}} \right)
\, \left( {{M_\odot}\over{M}} \right) \,
\, \left( {{10\, kpc}\over{D_S}} \right)
\, \left( {{1}\over{x\, (x-1)}} \right)
\end{equation}
Here we have assumed that there is a \pl\ located a distance $\mu\, R_E$
from the central star; the mass and luminosity of the central star are
$M$ and $L$, respectively. As usual, $D_S$ ($D_L$) is the distance to the 
lensed source (lens), and $x=D_L/D_S$. Given that (1) the conditions
that lead to life may be flexible, (2) the effects of radiation incident
from the star are likely to be strongly influenced (either enhanced or
diminished) by the \pl 's atmosphere, and (3) internal heating
from geological processes or radioactive materials may be important,
it is not clear how large a range of values of ${\cal F}/{\cal F}_\oplus$
may be compatible with the development of life. We will therefore simply
use ${\cal F}/{\cal F}_\oplus=1$ as a guideline.

\begin{figure}
\vspace{-1 true in}
\plotone{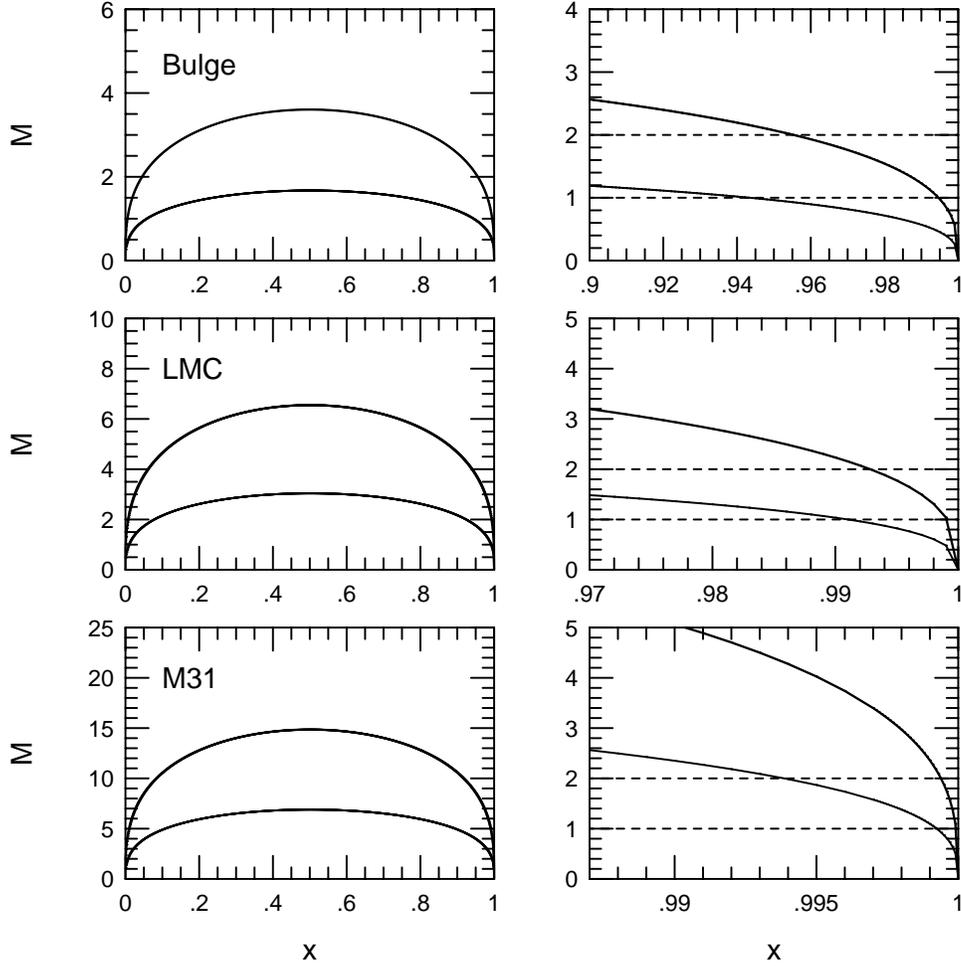}
\vspace{-2.0 true in}
\caption{Plotted is $M$ vs $x,$ where $M$ is the mass of the central star 
of the lens system and
$x=D_L/D_S.$ We have assumed that there is a \pl\ located at 
$a=\mu\, R_E,$ with $\mu=1.5$,
 and have imposed the condition: ${\cal F}/{\cal F}_\oplus=1.$
Upper curve in each panel: $L=M^{3.5}.$ Lower curve in each panel: 
$L=10 M^{3.5}.$ For the Bulge [Magellanic Clouds, M31], we have assumed
$D_S= 10$ kpc [$60$ kpc, $700$ kpc]. 
The plot in the upper left panel is for the Galactic Bulge;
the upper right panel is a blow-up of the region in $x$ that corresponds to
the lens, as well as the lensed source, being located in the Bulge.
Horizontal dotted lines are drawn at $M=M_\odot$ and $M=2\, M_\odot$ 
The left-right pairs in the middle and bottom panels show the same
quantities for the Magellanic Clouds and M31, respectively.}
\end{figure}

Figure 11 shows the relationship between $M$ and $x$ for those systems
that satisfy the relationship ${\cal F}/{\cal F}_\oplus=1.$ We have assumed
that $L=M^{3.5}$ in the upper plot, and $L=10\, M^{3.5}$ in the bottom plot;
the former would be appropriate for main-sequence stars, while 
the latter would be appropriate for slightly evolved stars.
As Eq.\ 18 and Figure 11 make clear, if the planets we will discover via
microlensing are to have an incident flux comparable to the flux
incident on Earth, their stars will generally 
(although not necessarily) 
be more luminous than our sun.
This has two obvious implications. The first is that the length of
time during which the planet would have this flux incident will be
shorter than the time to date that the Earth has had roughly this flux incident.
This is because the system's star may need to be more massive than the
Sun, or even slightly evolved. The time elapsed from the
formation of the star until the present time could range from
less than $0.1$ the present age of
the sun to times comparable to the sun's main-sequence lifetime.
We do not know how long it takes for complex life forms
to develop, but it may be that the process is fast
enough that intelligent life can develop and thrive during a time
significantly shorter than the present age of the Sun.
Indeed, it is likely that a long sequence of independent processes must
occur in order for intelligent life to develop; thus, the probability
distribution may well be log normal, and the likelihood of intelligent life
developing in times much shorter than the time apparently
taken
on Earth may be significant.

The second implication is that, since the central star must be
fairly luminous, it may contribute a non-negligible fraction
of the light incident along the line of sight to the lensed source; i.e.,
the light we receive may be strongly blended. We return to this issue in the next section,
but note here that it is potentially
 important, because when the \ps 's star emits
enough light to influence the lensing \lc\ in a detectable way,
we may be able to learn more about the \ps .  

We may also argue, as follows, that planets likely to be deemed
``Earth-like" have masses within a factor of $\sim 15$ of the mass
of the Earth.  Assuming that we would like a
rocky surface, we also assume that the planet's average density should be
similar to that of Earth.  This means that the acceleration due to gravity,
$g$, scales as the cubed root of the planet's mass, $m$.  Thus, $g$ will be
within a factor of 2.5 of $g_\oplus$ (= 9.8 m/s$^2$) if $m$ is within a factor
of 15 of $m_\oplus$.  Increasing or decreasing the value of $g$ will lead to
different atmospheric contents; for any given atmospheric temperature, there
is a lower limit to $m$ (hence $g$), below which an atmosphere will not be
retained. This means that, excluding finite-source size effects (which
can significantly increase the duration of  some events),
\ev\ durations for Earth-like \pl s are likely to range from $\sim 2$ to
$\sim 32$ hours.

We should also consider the possibility that \ml\ (whether by resonant
or wide \pl s) is most likely to discover the outer \pl s in a
system that may contain closer planets experiencing Earth-like
conditions. In this case, the \ml\ \ev s serve as beacons directing
us to the \ps\ the outer \pl s inhabit. Whether the \pl s that serve as
lenses themselves harbor life, or whether they merely serve as
bellwethers, it is important to consider whether additional observations
can help us to learn more about the \ps .

\section{Blending}

If the central star of a \ps\ that serves as a lens is fairly
luminous, then its light will blend with that from the
lensed star. 
Let $f$ represent the fraction of the baseline flux contributed
by the lensed star.
 We expect that in most cases,  
$f$ is a function of wavelength.           
If studies of the lensing \lc, or spectra taken during the \ev\ provide
evidence that light from the lensed star
is blended with light from other sources, 
 we can hope to measure the value of $f$ in each
of several wavebands. (See, e.g., \rd\ \& Esin 1995.)
 This information, combined with our
study of the baseline flux, allows us to obtain a spectral type for the
lensed source and also to study the spectral character of the 
unlensed light. If we are then able to determine that the unlensed 
light, or at least a quantifiable portion of it, emanates from the 
star in the lens system, we may then be able to
determine the spectral type and possibly the mass of
that star as well. 
If we are able to repeat this process for a
set of planetary-system \ev s, we  
will thus be able to gather statistics
about the types of stars which, in the distant reaches of
our Galaxy, or in other galaxies, have \ps s.
When the light curve allows us to determine the mass ratio $q$ between at
least one \pl\ and the central star, then the mass of that \pl\ is also
determined.  
Note that the process described above can be carried out 
even if the track of the source passes close to a \pl, but
 not close to the central star.

When the central star of the \ps\
 is luminous, particularly if its flux comes close to
satisfying the criterion studied in the last section
(${\cal F}/{\cal F}_\oplus \sim 1$), then $f$ can indeed be small enough for the
effects of blending to be measurable.
If, for example, the apparent $V$ magnitude of the combined light
coming along the line of sight from a lensing \ev\ is [$19.5, 17.0,
15.7$], then $f\leq 0.1$ if the lens is 
located in the Bulge and is a main-sequence star of mass
roughly equal to [$1, 2, 3$] $M_\odot$.
A small value of $f$ can allow us to reliably determine
the effects of blending and to thereby learn more about the lens.
Detection of the \ev\ can, however, become somewhat more difficult.
    
\subsection{The Effects of Blending on Event Detection}

When light from the lensed source is blended with light from other
sources, the observed \mage , $A_{obs},$ is smaller than the true
\mage , $A.$ 
\begin{equation} 
(A_{obs}-1)=f\, (A-1) 
\end{equation} 
Thus, in order for a \lc\ perturbation to be brought above the 
detection limit, $A$ must be larger than it would otherwise have to
be, the projected distance between the source and lens must be smaller,
and the \ev\ will consequently appear to have a shorter \du .
To describe this effect systematically, \rd\ \& Esin (1995)
introduced the ``blended Einstein radius", $R_{E,b}$.  
As before, let $A_{min}$ be the minimum peak \mage\ needed for \ev\ detection.
Here, however, 
we 
consistently define the \du\ of each \ev\ to be time during which the \mage\
was greater than $A_{min}.$ The expression for the blended Einstein radius
is then  
\begin{equation}
R_{E,b}= R_E
\sqrt{2}\sqrt{{{(A_{min}-1)+f}\over{\sqrt{(A_{min}-1)^2 
+ 2 (A_{min}-1)\ f}}}-1}.
\end{equation}
Figure 12 illustrates the influence of blending on \ev\ \du\
as a function of both $A_{min}$ and $f.$ 
For a given value of $f,$ the \ev\ \du\ is longer if $A_{min}$ is
smaller. Thus, increasing the photometric sensitivity to accomodate
smaller values of $A_{min}$
should increase the detection rate.
Note that if $A_{min}$ is $1.06,$ then even if $f=0.18,$ the
\ev\ \du\ (which would have been $2\, \tau_E$), is reduced by only a factor
of $2$ to $1\, \tau_E.$ 
Thus, while blending does tend to decrease the \ev\ \du, making more 
frequent \mo\ desirable, we can expect to be able to
detect a large majority of \ev s by using sensitive photometry.
We note that, if $A_{min}$ could be reduced to $1.02$--a formidable
task for a large-scale \mo\ program--then for $f=0.03,$
an \ev\ that would have lasted for $3\, \tau_E$ will have an observed
\du\ of $1\, \tau_E.$   
\footnote{Wambsganss (1997) has illustrated    
that, when there is a \pl\ in the \zres, evidence of the
\pl 's presence can be exhibited relatively early
during the rise of the underlying
stellar-lens \ev .  
In such a case, a smaller value of $A_{min}$ will also help us to
discover resonant \ev s that happen to be subject to blending.}

\begin{figure}
\plotone{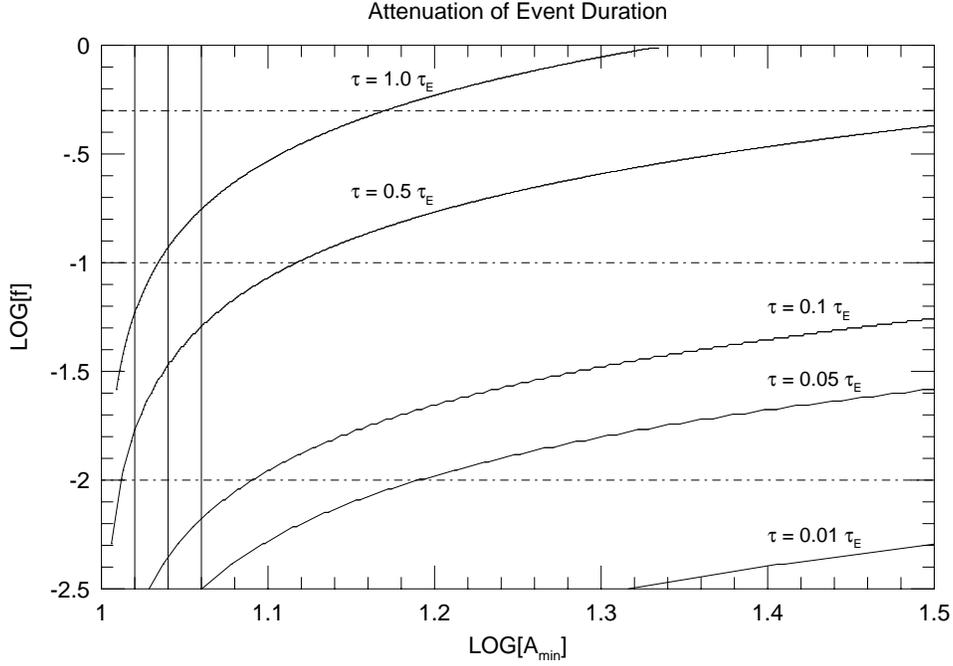}
\vspace {-4 true in}
\caption{
Log[f] vs Log[$A_{min}$]. 
Source tracks going straight through an Einstein diameter have been
considered. 
$\tau_E$ is the time needed for the source to cross a distance
equal to the Einstein diameter.  
The observed \ev\ \du , $\tau,$ is defined to be the time during which
$A>A_{min}$.  
Because each \ev\ was subject to blending ($f < 1$),
$\tau$
 is attenuated relative to the event duration
without blending. 
Note that, for $A_{min} < 1.34,$ the duration of the observed \ev\ would be
greater than $\tau_E$, were there no blending.
Each curve shown corresponds to a fixed ratio $\tau/\tau_E.$
The horizontal lines correspond to $f_V = 0.5, 0.1, 0.01,$ proceeding
from top to bottom. For reference, we note that if the lens is a main-sequence
star with mass [$1, 2, 3$] $M_\odot,$ then $f_V=0.5$ if the total
apparent V magnitude along the line of sight is [$20.2, 17.6, 16.4$];
similarly, $f_V=0.01,$ for $M_V$ = [$19.4, 16.9, 15.6$].       
Vertical lines correspond to $A_{min}=1.02, 1.04, 1.06;$ these 
provide a comparison which  
demonstrates the improvement in \ev\ detection relative to the case $A=1.34.$
}
\end{figure}
\subsection{Studying Blending}

As long as the presence of a blend of light from sources other
than the lensed star does not
prevent us from detecting or identifying an event, blending can be a
great boon to the analysis of \ml\ \ev s.
Using blending to full effect requires several steps.

\noindent (1) {\sl Spectra:}\ Spectra taken during the \ev\
can help us to determine the spectral type of the lensed source.
During the event,
the lensed source contributes a fraction, $f_{ev}(t)$, of the
incident light.
\begin{equation}
f_{ev}(t)={{f+(A_{obs}(t)-1)}\over{1+(A_{obs}(t)-1)}}
\end{equation}
Thus, spectra taken at peak can be compared with spectra taken
at baseline (and possibly at other times during the \ev ) to
determine the spectral type of the lensed star, and to thereby
better quantify the contributions of any unlensed source of
light as a function of wavelength.

\noindent (2) {\sl Light Curve Fitting:}\ If the \lc\ is observed
in several wavebands, then the fit to the \lc\ will determine
$f$ in each waveband separately. This information, combined with
the baseline magnitudes, also provides a way of determining the
spectral type of the lensed star, and therefore the spectral characteristics
of the unlensed light. Information about blending drawn 
from studying the \lc, should confirm and complement information
drawn from spectral studies. 
 
\noindent (3) {\sl High-Resolution Follow-Up:}\ The determination
that there is light incident from an unlensed source does not
necessarily indicate that the extra light is emanating from the lens
itself. In fact, the fields studied by the monitoring teams are
crowded, so there is a high probability that another star, independent
of both the lens and lensed source, may be contributing to the incident
flux. High-resolution observations can determine whether the ``extra" light
is emanating from one or more stars, distinct from the lensed source,
that were not resolved by the \mo\ teams. The density in
the Bulge, for example,
is small enough that it is 
unlikely that independent stars 
located along the direction to the \ev\ 
(apart from the lens),  
will not be resolved by HST.
Because the angular separation between the lens and source is on the 
order of milliarcseconds, we cannot, however, resolve the    
lens and source immediately after the \ev , 
even if the lens system contains a luminous star.
Thus, the light from each star resolved by high-resolution follow-up
can be subtracted (as a function of wavelength) from the baseline flux
to determine how much light, in each of several wavebands, may 
be emanating from the lens system itself.  
After a time determined by $v_t,$ we should be able to
resolve the lens system and test the hypothesis that the lens provides
the light unaccounted for by other resolved stars;
the wait-time is typically on the order of decades. Until the lens
and source are resolved, we cannot be certain that the additional light 
does come from the lens system.
It is after all possible that yet another star has a small angular
separation from both  
lens and lensed source; the likelihood of such a coincidence can be
evaluated.  
In many individual cases, and certainly for a statistical sample of \ev s,
we expect that the
most likely alternative is that the additional light emanates from the
lens system.    
Spectral and high-resolution measurements may provide
enough information to determine the spectral
type of the central star of the \ps .
Although the distance from the lens will not be known,  
we will, 
particularly with the input 
from the lensing \ev\ itself, 
be able to place limits 
on its mass. (See Udalski {\it et al.} [1994] 
for a discussion of this problem for the OGLE 7 \ev.) 
The mass of the central star, together with the determination of $q$
for any \pl\ lens, allows us to compute the \pl 's mass.

\noindent (4) {\sl High-Spectral-Resolution Follow-Up:}\ 
We note that Keck HIRES observations
have produced a high-resolution spectrum for a star (with $M_V=17$)
 in Baade's window,
allowing a  detailed abundance analysis
to be carried out (Castro {\it et al.} 1996). 
This would also be possible for stars in the \ps s we discover
via \ml . 
Ideally, one would like to go further, and to
 achieve spectral resolution good enough
to detect evidence of low-velocity motion, which may
indicate the presence of \pl s closer to the \cs\ 
than the \pl s discovered via \ml . Extrapolating from the 
HIRES work and from more general considerations about radial velocity
measurements (Latham 1996), we estimate that 
the photon flux we are likely to receive from the lens system is
too small (generally by more than 2 orders of magnitude) 
for the present generation
of telescopes and detectors to be effective. 
Thus, although possible in principle,  such fine radial velocity
 measurements  
are presently beyond our reach for lens-system stars in the Bulge.

\section{Finite-Source-Size Effects} 

Finite-source-size effects are known to be important when attempting to
detect evidence of a low-mass (e.g., Earth-mass) \pl\ located in the \zres;
in this case, it is the comparison between the size of the caustic structures 
and the radius of the source as projected onto the lens plane,
that is important. The caustic structure, however,
shrinks and therefore plays a
less important role as the separation between the star and \pl\
increases.  Thus, for \pl s in \wo s, the important comparison is
between the size of the \pl 's \er\ and the size of the source as
projected onto the lens plane. 
\footnote{The caustic structure may play a more important role in
a small fraction of overlap \ev s than it does for wide-orbit \ev s 
generally.}
Until now we have assumed that the size of the lensed source can be
neglected 
 in comparison to the size of the Einstein ring. 
For lensing by \ps s, 
this is not always
a good assumption.

Let $r$ represent the ratio between the radius of the source star and the
Einstein radius of the lens.  
$r=R_S/R_{E,i},$ with $R_S$ the radius of the lensed source
as projected onto the lens plane. 
\begin{equation}
r=5.2\times 10^{-3}\, \left( {{R_S}\over{R_\odot}} \right)
     \left[ \left( {{M_\odot}\over{M}} \right) \,
     \left( {{10\, kpc}\over{D_S}} \right)
    \, \left( {{1}\over{x\, [1-x]}} \right) \right] ^{{1}\over{2}},
\end{equation}
\noindent 
where $M$ is the
mass of the lens. If the lens has the mass of Jupiter
and if the source and lens are both in the Bulge ($D_S = 10$ kpc, $x=0.9$),
then $r\sim 0.05.$ Witt \& Mao (1992) found that the finite size of
the source begins to influence the shape of the \lc\ when $r\sim 0.1.$
Since typical source stars are likely to be larger than $R_\odot$,
 finite source size effects may play some role even when a
\pl\ as massive as Jupiter serves as a lens. The significance of their role
increases as the mass of the \pl\ lens decreases.

\subsection{Finite Source Size and Detectability} 

The \lc\ changes in two ways when the 
size of the source cannot be neglected. First, whereas the \mage\ of
a point source lensed by a point mass can become arbitrarily large when
the distance between the lens and projected source position becomes
arbitrarily small, 
the \mage\ has an upper limit  
when the source size is finite. 
Thus, the peak of the \lc\ becomes attenuated. 
The larger the source size relative to the \er, the more
pronounced the attenuation.
The maximum magnification possible
for a disk of constant surface brightness, for example, is
\begin{equation}
A_{max}={{\sqrt{R_S^2+4}}\over{R_S}},
\end{equation}
This formula indicates that the
size of the source must be fairly large in order for the
peak \mage\ to be brought below the level of detectability, 
particularly if the photometric
sensitivity is good.
For example, a magnification of $1.34$ can be achieved
even if the size of the source is $\sim 2.25 R_E$,
and a magnification of $1.02$ can be achieved
even if the size of the source is $\sim 10 R_E.$ Note that 
a projected source size of $10 R_E$
corresponds to a $\sim 200 R_\odot$ ($20 R_\odot$) stellar radius
when the lens is of Jupiter-mass (Earth-mass). (We have assumed $x \geq 0.9$.)
Thus, 
in spite of the decrease in peak \mage, a goodly fraction
of stars in the source population can, when are lensed by \pl s in \wo s,
produce \ev s that should be detectable, at least with good photometry.

The second effect of finite source size on the \lc\ is to broaden
the width of the perturbed region of the light curve, since some portions
of the source may be significantly magnified well before the center of the 
source 
achieves its closest approach to the lens.  
The \lc s can become almost flat-topped.
Thus, magnifications close to the peak value may be sustained during
the time it takes the source to travel several Einstein diameters,
and the event appears to
last longer than it would have had the lensed source been a point source.  
This can be a boon to the detectability of \ev s due to
planet-mass lenses, since one of the greatest barriers to detection is
the short lifetime of the \ev . 

\begin{figure}
\vspace {-1 true in}
\plotone{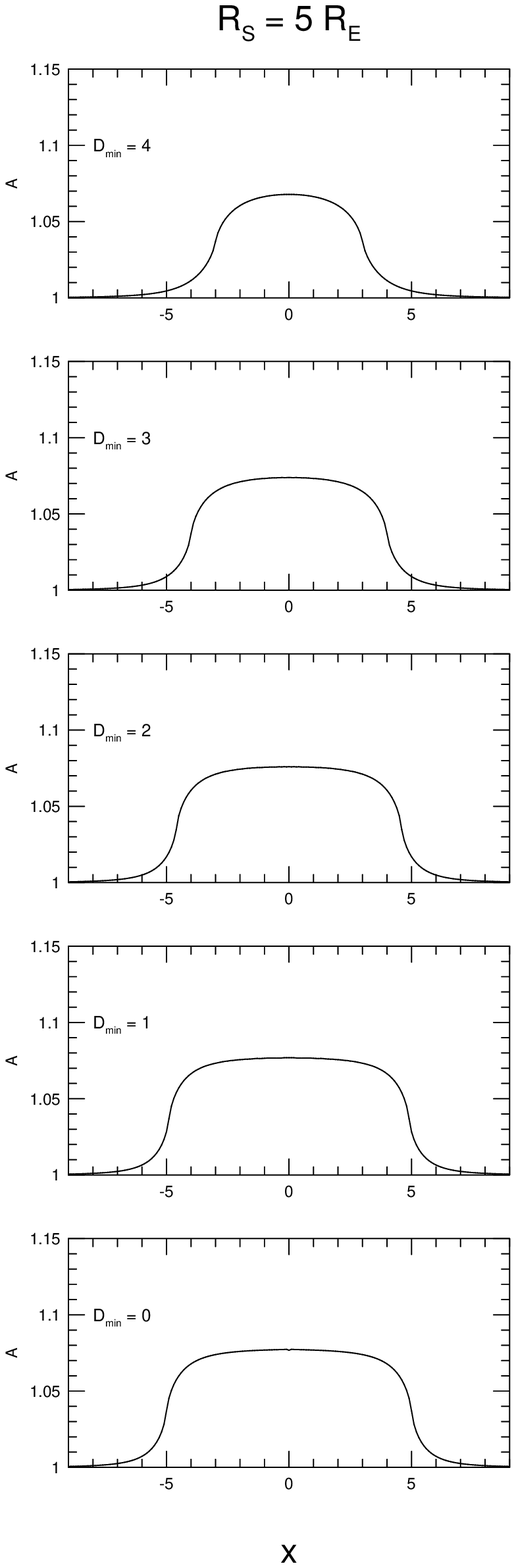}
\vspace {-.5 true in}
\caption{
In each panel, the projected size of the source in the lens plane is
$5\, R_{E,i}.$ The distance of closest approach, $D_{min}$, decreases from
$4\, R_{E,i}$ in the upper panel to the $0$ in the lower panel.
Note that the peak \mage\ is larger than $1.06$ in all cases.
Furthermore, if the measured
\ev\ \du\ is the time during which the \mage\ is larger than $1.06,$
then
these \ev s last up to   $2.5$ 
(bottom panel) times as long as they
would have, had the lensed source been a point source. 
}
\end{figure}

\begin{figure}
\plotone{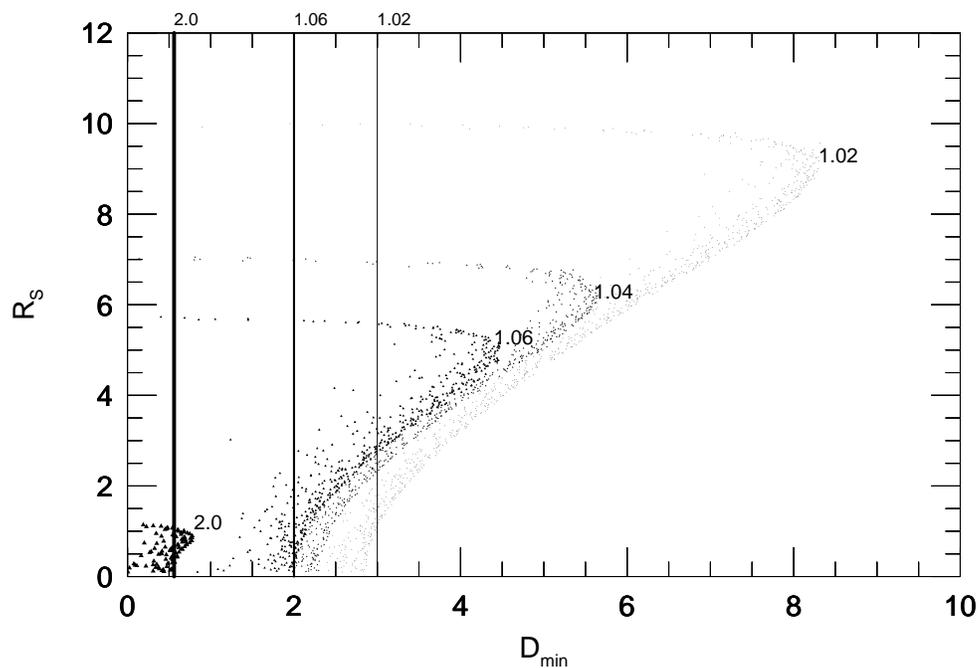}
\vspace {-4 true in}
\caption{
The source radius, $R_S,$ is
plotted against $D_{min},$ the distance of closest approach.
Both quantities are expressed in units of the Einstein ring radius.  
Each point
corresponds to an event in which the peak \mage\ was roughly equal to  
the value listed on the upper right of each curve. 
For each value of $R_S,$ only the points with the largest values of
$D_{min}$ leading to the listed \mage\ are shown. 
We have assumed that the disk of the lensed source has constant surface
brightness.
The value of $D_{min}$ for a point source is shown, for $A_{min}=
1.02, 1.06,$ and $2.0$ [vertical lines]. 
}  
\end{figure} 

\begin{figure}
\vspace {-1 true in}
\plotone{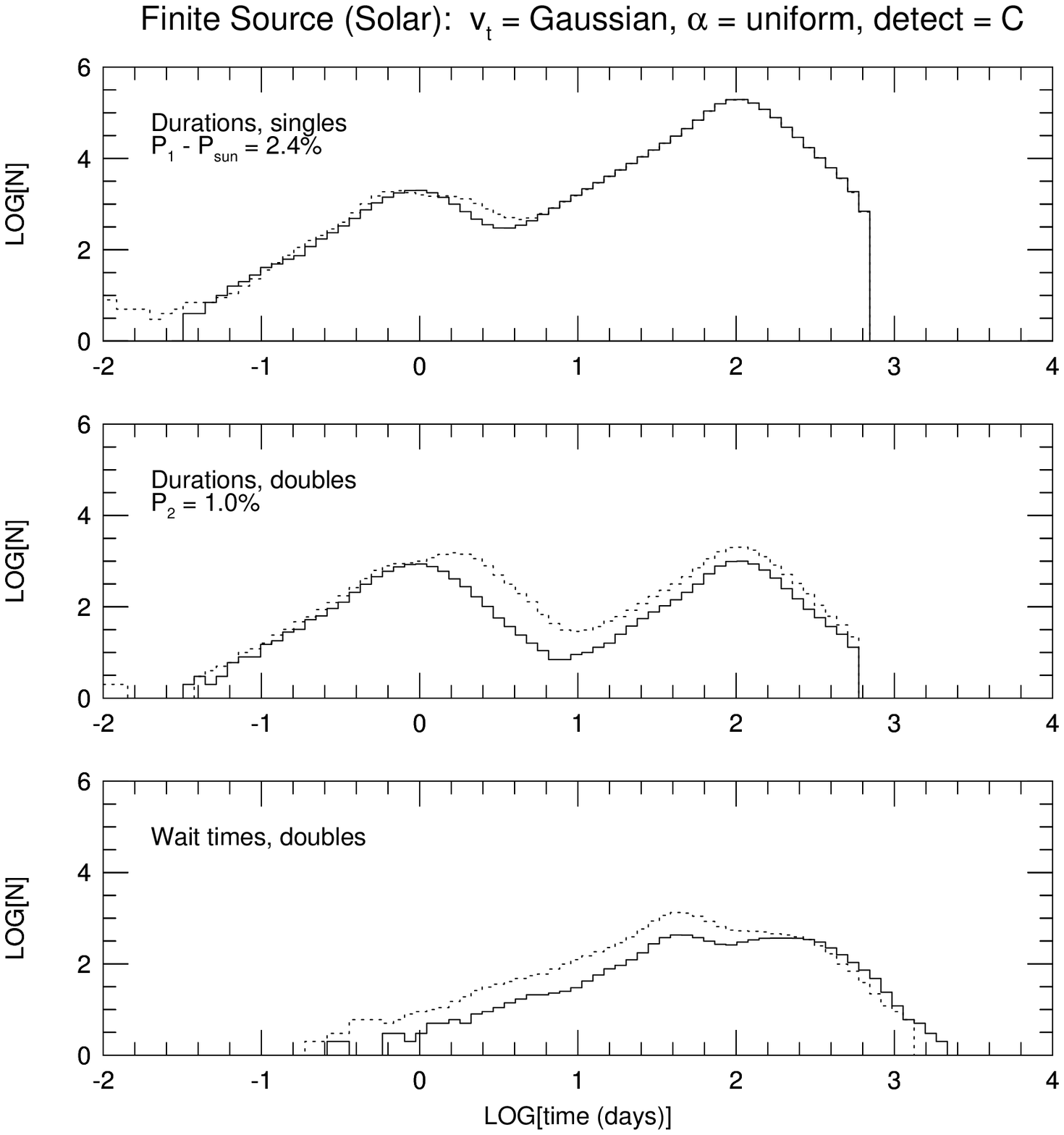}
\vspace {-2 true in}
\caption{Results of a 
Monte Carlo simulation 
that includes finite-source-size effects for a lensing system with
9 Earth-mass planets whose separations from the
solar-mass central star are the same as
those of the planets in our solar system.  
We have assumed that the source radii are distributed  
uniformly from $3\, R_E$ to $10\, R_E.$
In addition, we assume that lensing by Earth-mass \pl s cannot be detected if
$R_S > 7\, R_E.$ Thus, roughly $43\%$ of the sources could not produce
identifiable lensing \ev s. For the remainder of the sources, we used
the approximation that $D_{min}\sim R_S.$ (See Figure 14.)
In these simulations we used: $v_t$ = Gaussian, $\alpha$ = uniform, detect = C.
For comparison,   
the distributions 
for the equivalent simulation of
lensing by the model of the  solar system used in \S 4,
with the sources taken to be point-like,
 are shown by the dotted curves.
}
\end{figure}

Figure 13 displays a sequence of \lc s for the case $R_S = 5\, R_E.$
When comparing the different \lc s, note that the peak \mage\
does not change much, even as the distance of closest approach, $D_{min},$
changes from $0$ to $4\, R_E$. If the photometric sensitivity of the 
observations was such that a \ev s with peak \mage\ above $1.06$
could be detected, then all of these \ev s would be detected.
The primary difference as $D_{min}$ increases, is that the time \du\
of the observed \ev\ decreases. Even so, for $D_{min}$ as large as $4\, R_E,$ 
the 
observed duration
is still roughly equal to the time a point-source \lc\ would remain above
$A= 1.06.$ 
The relationship between $D_{min}$ and $R_S$ is explored more systematically in 
Figure 13. 
The implications
are that:
(1) depending on the distribution of source sizes,
the \dtn\ probabilities can be increased by \fsse , and (2) \ev s
that can be detected can also generally be monitored for longer periods of
time. 

Gould (1994) was able to find point-source fits
to \lc s associated with large sources. This degeneracy of the 
\lc\ shape, however, need not prevent the identification of
individual \ev s as having been influenced by \fsse , since 
spectral information can break the degeneracy. Real stellar
disks exhibit brightness profiles and spectra that have spatial
structure. When a star for which $r$ is not negligible is lensed,
the observed spectrum will therefore be time-dependent, and 
can be used to study the lensed star. 
(See, e.g., Witt \& Mao 1994,
Witt 1995, Gould 1994, Loeb \& Sasselov 1995,
Simmons, Willis \& Newsam 1995, 
Sasselov 1997, Heyrovsky \& Loeb 1997, and, for an observation, 
Alcock {\it et al.} 1997.)
This time dependence also helps to confirm the interpretation of the
\ev\ as due to \ml\ of a source whose size has influenced the \lc .
Indeed, when the source is large relative to the Einsein ring,
it is as if a magnifying glass were scanning the face of the lensed star, 
thus providing a good deal of information
about the source star and about the \ev . Spectral studies, even during 
\shdn\ \ev s  can therefore be valuable
in breaking the degeneracy for individual \ev s.  Even if spectral information is not 
available, however, statistical analysis of the ensemble of
\ev s would be able to indicate that \fsse\ had played a role.
In particular, the relationship between peak magnitude and duration
would be different than it would be for a set of true Paczy\'nski \lc s;
for large sources, the peak magnification is closely related to
source size, while the duration continues to have a more direct 
relationship to the 
\dca .     

To better quantify the influence of finite source size, 
we have carried out simulations using the solar system model
and a power-of-3 model to study the detectability of Earth-mass
\pl s. These results are summarized in Table 4 (\S 10) and are illustrated
for the solar system model in Figure 15. (The input assumptions are described
in the caption to Figure 15.) 
Because their Einstein rings are smaller, Earth-mass \pl s are expected
to yield a smaller \ev\ rate. The rate would be 
smaller than that due to Jupiter-mass \pl s by a factor of
roughly $18$. Yet, in the power-of-3 model, where all of the \pl s in our
original model (\S 4) had a mass equal to that of Jupiter, 
the computed attenuation of the
Earth-mass system relative to the 
Jupiter-mass system is only a factor of $\sim 3.$  Even though
\fsse\ made \ev\ detection
impossible when the Earth-mass \pl s lensed roughly $43\%$ of the 
sources, the fact that the detection rate was higher for the remainder
of the sources provided a net increase in the detection rate over what might
otherwise have been expected. The relative increase is even more
pronounced for the solar-system model, since Saturn, Uranus, Neptune, and
Pluto are responsible for many of the isolated \ev s when a model
of the solar system (\S 4) serves as a lens, and these each
have a smaller Einstein radius 
 than Jupiter.
Although more
realistic stellar surface brightness profiles should be used,
the basic result that Earth-mass \pl s are detectable should be robust.  
Also robust is the increased \ev\ \du\ for a range of values of
$R_S$. For a fixed value of $A_{min},$ this corresponds to
a larger value of the effective width, $w_i$, and therefore a larger
value of $n=w_i/R_{E.i}$.  This can increase the detection probabilities
above what is expected if the source is point-like,
as predicted 
 and as demonstrated by the simulations. 
 
\subsection{Planetary Masses and Finite-Source-Size Effects} 

When spectroscopic studies of the lensed source
can establish its real physical size, finite-source-size effects
may then allow us to derive the true size of the Einstein ring.
This means that,
if the value of $r$ can be measured from the data,
 $v_t$ can be determined. Furthermore the degeneracy
in the lens mass associated with the Paczy\'nski \lc\
is partially broken, since knowing the value of $R_E,$ and, through
spectral studies, the value of $D_S,$ allows us to 
infer
\begin{equation}
M\, x\, (1-x)=C,
\end{equation}
\noindent where $C$ is a constant whose value is measured.
Given the distribution of likely values of $x$ (from $0.9$ to $0.99$,
for example, in the Bulge) this generally
constrains $M$ to within a factor of $\sim 5.$ Thus, the
hypothesis that any given \ev\ is due to lensing by a  planetary-mass
can be meaningfully tested.

\section{Moons, Asteroids, and Comets}

Expanding the \ml\ searches  for \pl s to those in \wo s
allows the presence of each wide-orbit \pl\ to contribute to the
probability of detection, and also allows for possibly
complex signals, containing information about the masses and spatial
distribution of several \pl s in the
\ps.
It is therefore interesting to consider that the complexity of \ps s
extends beyond the presence of multiple \pl s, to the possible
existence of moons revolving about the \pl s, and also of belts or rings of
space debris, such as asteroids and comets.
What is the probability of detecting these features through \ml\ studies?

\subsection{Moons}

The \ml\ detection of a moon about a \pl\ requires that
(1) the \pl\ about which it revolves either be in the \zres\ or in a wide
orbit,
(2) the moon be in
either a resonant or wide orbit about the \pl,
(3) the duration of the perturbation
due to the moon is long-lived enough to be detectable, and (4)
\fsse\ do not wash out the
signature of the moon's presence.

\subsubsection{Planets in the Zone for Resonant Lensing} 

If the \pl\ is in the \zres , then the presence of the moon
will typically complicate the caustic structure and the associated pattern
of the \lc . For \pl s in \wo s, the presence of a moon in
the \zres\ about the \pl\ could mean that
the \lc\ associated with lensing by the \pl\ is more likely to
exhibit the behavior associated with caustic crossings. 
The size of the caustic
structures due to the presence of a moon will generally be small;
consequently the associated perturbations may be difficult to detect. 

\subsubsection{Wide Orbits}
 
If the moon is in a  \wo\ about a \pl\ which is itself in a \wo\ about
its star, it is relatively straightforward
to determine what the signature of the moon's
presence is, and to assess its detectability.
Going through the $4$ criteria listed above, we 
note that the first two conditions are demonstrated 
to be feasible by the example of our Solar System. Were
an identical system to be placed in the Bulge, Saturn and Neptune
would both be in \wo s, with projected separations significantly larger than
the sun's Einstein radius most of the time and for most system orientations.
Pheobe orbits Saturn in an orbit that is wide
with respect to Saturn's Einstein radius; similarly, Nereid orbits
Neptune in a wide orbit (Norton 1989).
With regard to criterion $3$, if we would like an \ev\ due to
lensing by a moon to last
at least $1$ hour, then, assuming that an \ev\ due to the central
star lasts $100$ days, the mass ratio between the moon and the star
can be as small as $1.7 \times 10^{-7}.$ If the central star has a mass
equal to that of the sun, the smallest detectable moon could have a mass as small as
$m_{min}=3.5 \times 10^{26}$ gm. 
This is roughly $5$ times the mass of our own moon.

Criterion $4$, the limit placed by finite-source-size, 
is also important. 
If we impose the condition that $r$ have a maximum value of [$10, 5, 1$], then
we find that the minimum moon mass has a value of [$5\times 10^{25}$ gm,
$2\times 10^{26}$ gm,  $5\times 10^{27}$ gm]. These simple calculations
make it clear that \fsse\ will be important in the \ml\ study of moons.
Those moons that do give rise to detectable \ml\ \ev s
may yield \lc s whose shape is significantly influenced by \fsse .
In these cases, the effects can be beneficial, since the \ev\ durations
may be 
prolonged by \fsse, lasting several hours.

We note that, if its density were $6$ gm/cm$^3,$ 
a moon of mass equal to $3.5\times 10^{26}$ gm would have a
radius equal to $\sim 2400$ km. 
In fact, the density of the wide moons in our own solar system is estimated
to be smaller than $6$ gm/cm$^3,$ while their physical
radii are smaller than $1000$ km (Norton 1989). 
It is therefore unlikely that \ml\
observations of our solar system would reveal the presence of its many moons.
There does not seem to be a fundamental reason, however, why other
systems might not have moons that can be detected via \ml .
Thus, it may be productive to institute even more frequent
\mo\ just after a \shdn\ \ev; if the \ev\ is due to
a \pl, careful frequent \mo\ after it has ceased 
would improve our ability to 
detect any associated moons.

\subsection{Asteroids and Comets}

It is also interesting to consider the detectability of a belt of debris,
such as the solar system's asteroid belt, or a cloud of comets.
Each isolated member of such a belt
is likely to be much too small to be detected through its action as an 
individual lens. It may be
possible, however, that the projected mass surface density is close to the
critical value, $\sigma_{crit}$. 
In this case, the track of the source could pass across
a caustic structure, giving rise to 
variability in the \lc. (See also Keeton \& Kochanek 1997.) The 
time scale of the 
perturbations would be set by the time required for
the track of the source to pass through the belt.

The critical density is
\begin{equation}
\sigma_{crit} = {{c^2}\over{4\, G\, D_S \, x\, (1-x)}} =
\Bigg(1.1\times 10^5\, {{gm}\over{cm^2}}\Bigg)\, 
\Bigg({{10\, kpc}\over{D_S}}\Bigg)\,
\Bigg({{1}\over{x\, (1-x)}}\Bigg) 
\end{equation}
This surface mass density would not be achieved 
by the belts of asteroids and comets associated with our own solar
system, unless $D_S$ is much larger than the distance to the Bulge, and the
value of $x$ is
neither too large nor too small.

Consider a belt of debris of total mass $10^{30}$ gm,
with a projected extent of $0.01$ AU by $1$ AU. The surface density
would be $\sim 4.4 \times 10^5$ gm/cm$^2$. With $D_S$ equal to $10$ ($60$)
kpc and $x=0.5$, we derive $\sigma_{crit} = \sim 4.4 \times 10^5$
($\sim 7.3 \times 10^4$ gm/cm$^2$). Thus, it is not inconceivable that,
early in the evolution of a \ps , $\sigma_{crit}$ can be achieved,
leading to interesting signals, particularly for favorable system orientation
relative to the line of sight.

\section{Detection Strategies}

The strategies best suited to the discovery of wide planetary systems have
much in common with those best suited to the discovery of planets in resonant
orbits.  In both cases it is important to have frequent time sampling of
events or event features whose duration may be on the order of hours.  In both
cases, good photometry is important if we are to carry out meaningful tests of
the planetary-system lens hypothesis.  Thus, since a detailed search mechanism
is already in place to discover planets in the \zres, it is useful to ask how
effective that mechanism will be at discovering wide planets, and whether
slight modifications of it will make it more effective.

\subsection{Past and Ongoing Observations}

\subsubsection{\sl Short-\du\ \ev s}

 The monitoring teams
have discovered some short-duration \ev s, with the shortest
\ev\ on record lasting approximately $2$ days. 
To better understand what is known so far, we have considered \ev s along the
direction to the Bulge, since the MACHO team has already accumulated a store of
over $150$ Bulge \ev s about which some information is publicly available.
Specifically, we have used their paper on $45$ Bulge \ev s monitored in 
1993 (Alcock {\it et al.} 1997a),
and their ``alert" web pages (http://darkstar.astro.washington.edu),
which list approximate \du s for many of the 136 Bulge \ev s observed
during 1995, 1996, and 1997. Altogether we found $148$ \ev s  
(a) which  are apparently 
due to lensing by a point-mass, and  (b) 
for which approximate \du s are available. (We note however, that
the \du s posted on the web site come from fits that may be refined in the future.)   
The \dbn\ of \ev\ \du s is plotted in Figure 16. Note the appearance of
the \shdn\ peak. 
In addition, at least one \shdn\ \ev\ is not included in this graph;
96-BLG-9 is simply listed as a ``short-time-scale \ev", with no \du\ given.

\begin{figure}
\plotone{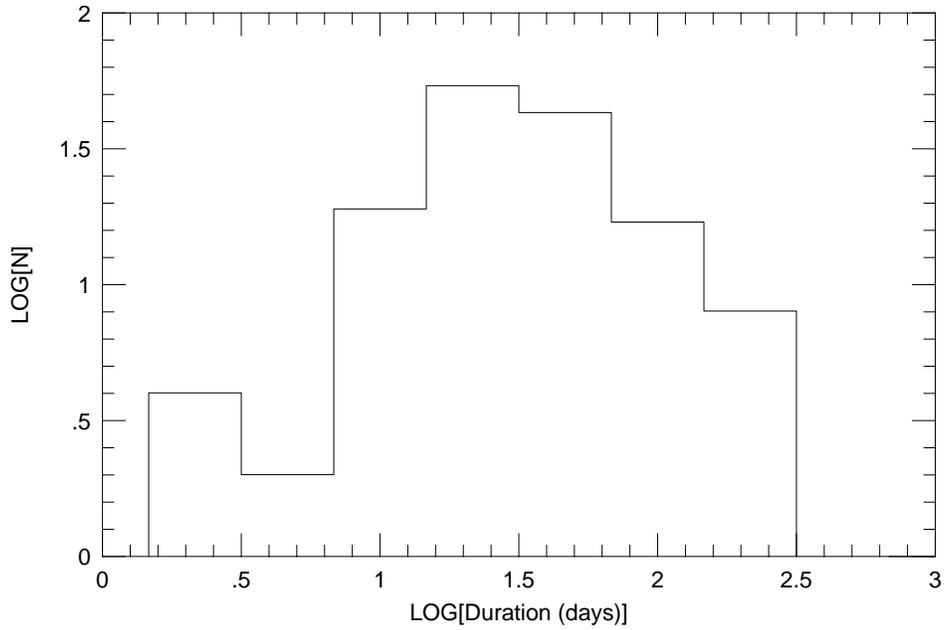}
\vspace {-4 true in}
\caption{
Distribution of \ev\ \du s for the MACHO team's set of Bulge \ev s.
Included is data from 1993 (Alcock {\it et al} 1997a), and the ``alert"
web page data for \ev s which occurred in 1995, 1996, and 1997.
All 148 \ev s which were not listed as binary \ev s, and for which
estimates of \ev\ \du s had been, made are included. 
}

\end{figure}

In light of Figure 16 we may ask whether the MACHO team has already
begun to discover concrete evidence for a signature due to \pl s in the
Bulge. This may be the case. (In fact 
Bennett {\it et al.} [1996] have conjectured that one short \ev\ may indeed
be due to a \pl\ in a \wo\ .) 
We emphasize, however, that a good deal of
further work would be required to clearly establish it. In fact, the
remainder of this section and the next are devoted to detailed discussions
of productive lines for further research. 
One certain conclusion we can draw from the MACHO Bulge data is that
the team has proved that it has the ability to detect \shdn\ \ev s.  
This is encouraging, since 
the detection strategies presently used are not optimized for the
discovery of 
short time-scale \ev s. 
\footnote{
This is an appropriate place to note that, while our choice of $1$ day
for the minimum time-duration necessary to detect a first \ec\ (detection
criteria A and B) is perhaps too optimistic to describe the present
search strategy,  
the fact that the \mo\ teams have been able to identify 
\ev s that last for $\sim 2$ days indicates that it may not be overly
optimistic.   
It can clearly be achieved in some fields even at present; 
were more emphasis to be placed on the discovery of short-time-scale 
\ev s, then it might be achievable for a larger portion of the data sets.
In addition, as indicated earlier, even if frequent \mo\ of a \shdn\
\ev\ does not begin in time to provide much more information about its
\lc, it could nevertheless  play a crucial role in discovering any
later repetitions.}  
Thus, the discoveries to date indicate that
the number of short duration events is large enough that a more
concentrated effort to find them is likely to bear fruit.  

In addition to being able to discover \shdn\ \ev s, the MACHO and EROS
teams have also been able to place upper limits on the numbers of such
\ev s along the direction of the LMC (Renault {\it et al.} 1997, 
Ansari {it et al.} 1996; Alcock {\it et al.} 1996).  

The follow-up teams can only study short-\du\ \ev s discovered by
the monitoring teams if the \ev s last more than a day. In fact,
the time presently taken to call an alert
may vary from $1$ day up to $\sim 5$ days (Axelrod 1997).
Thus, the chances that the follow-up teams will be able to
catch lensing \ev s due to \pl s with mass less than that of Jupiter are
small.
They may, however, be able to follow-up on
\ev s due to Jupiter-mass \pl s and larger, or on \ev s 
due to less massive \pl s if the \du s of the \ev s are extended by \fsse.

\subsubsection{Repeating events}
The early folklore on repeating events said that they should be
discarded: the microlensing of light from a specific star is such
a rare event that, if a \ml-like signal is seen twice when monitoring the
same star, the repetition is almost surely a sign that the observed variation is
due to
something other than \ml .
Although it has been
pointed out that binary sources (Griest \& Hu 1992) and binary lenses (\rd\ \& Mao)  
can each lead to \ml\ \ev s that repeat, it is nevertheless the case that
 one of the cuts used to eliminate
lensing candidates from further consideration is based on whether
the event appears to repeat. {\footnote {Of course the complication of
ensuring that a repetition is not due to stellar variability is serious,
since variability is much more common than \ml. 
Nevertheless, if a  star that has
never been observed to vary from baseline produces two deviations well-fit
by \ml\ models, it is certainly worth exploring the possibility
that the deviations are indeed due to \ml . In some cases blending or
finite-source-size effects may make it possible to carry out further
tests of the conjecture that multiple disturbances are due to \ml.}} 
   As a result, the \mo\ teams may not pursue the
implications of evidence of
repeating \ev s in their data sets.
We note that one ``repeater" is mentioned on the MACHO alert web site,
but the posted information 
does not indicate whether \ml\ seems a likely explanation for either 
perturbation.  

The follow-up teams do not, at present, continue to follow \ev s once the
detected flux has returned to its baseline value.

\subsection{Useful Modifications}

\subsubsection{Improved Measurements for both Resonant and Wide Planets}

Since detection of blending and finite-source-size effects
could
possibly clinch the \pl-\ml\ interpretation, it is worthwhile to 
search for evidence of these effects in every \ev . The analysis
of light curve shape
can play an important role; this analysis 
can be done after the \ev\ has ceased.
During the \ev , it would be valuable to obtain spectra,
particularly near the time of peak \mage. 
 
\subsubsection{Improved Measurements for Wide Planets}

The main difference between the resonant and wide cases is whether or not
there is a trigger that indicates  that frequent \mo\ should begin.
In the resonant case, an ongoing stellar 
lens \ev\ provides a convenient trigger. As soon
as any \ml\ \ev\ is identified, intensive monitoring that lasts for the 
\du\ of the \ev\ can begin. In the wide case, many of the \ev s that could
be better understood through intensive \mo\ 
have no convenient
trigger. This is particularly true for any \ev\ in which a \pl\
is the first or only lens. 
Thus, in order to optimize our ability to detect \pl-lens \ev s,
regular frequent \mo\ is called for. The optimal frequency however,
depends on $A_{min},$ the minimum peak value of the \mage\ 
for which an \ev\ can be reliably identified. 
This dependence is best expressed in terms of $n=w_i/R_{E,i}$. 
If $k$ is the number of repetitions exhibited by the \lc, the 
detection rate scales as $n^{k+1}.$ Not only are more \ev s discovered
when $n$ is greater than unity, but the \du\ of detected \ev s is also
longer, making less frequent \mo\ possible. 
Note that \fsse\ can increase the value of $w_i$ (hence $n$), even 
for a fixed value of $A_{min}$.  

The bottom line is therefore 
that 
good photometry is the key to increasing the detection rate of \pl s
in \wo s. Monitoring that is consistently more frequent
than that presently  carried out by the \mo\ teams is
also important, although hourly \mo\ may not be needed even for the 
detection of Earth-mass \pl s.

The design of the observing programs of the future should take these
requirements seriously. There are likely to be several
 different ways to accomplish them. For example, a system of \mo\
could achieve a detection rate $\sim 10$ times as high as the present system
does by (1) using photometry as good as or better than the present follow-up
teams, (2) using pixel techniques to identify lensing of stars that are
otherwise (i.e., at baseline)
 below the detection limit, and (3) \mo\ frequently.
This would mean discovering ${\cal O}(1000)$ Bulge \ev s per year. Not only 
would the numbers of \ev s be high, allowing us to detect some very
low-probability \ev s (such as multiple repetitions), but the quality of the 
data would be so high that we would understand each \ev\ and the population
of lenses and sources much better than we do at present. 
Another approach might be to have a world-wide network of telescopes, each   
taking deep images of a few fields once or twice per night. The detection
rate per field and the quality of the data would be high, but the number of 
\ev s discovered per year would be smaller, since fewer stars are being 
monitored. 
The total number of \ev s discovered per year 
might be comparable to the discovery rate of the 
present MACHO team.

Below we focus on modifications of the present strategies that would
increase the detection rate of \pl s in \wo s.
 
\subsubsection{Useful Modifications: The Monitoring Teams}

(1) {\sl Spike Analysis:} \
Along the direction to the Bulge it seems likely
that the large majority of the \ev s already
detected are due to lensing by ordinary stars. If a significant
number of stars have planets in \wo s, then short \ev s due
to lensing by these \pl s are expected.
A spike analysis of existing data from the Bulge could be
productive in either discovering evidence of short \ev s 
(in addition to those already discovered) or placing upper
limits on the number of short-time-scale \ev s that might have occurred.
As more data is collected by a larger number of \mo\ teams, systematic spike
analyses should become standard.

(2) {\sl Software Algorithms:}\
It is important to search the data sets for repetitions, perhaps in
conjunction with a spike analysis.

(3) {\sl Frequent Monitoring of Some Fields:}\   
The importance of the quest for short time-scale
\ev s suggests that it might be worthwhile for the teams to
each choose one or two fields which they attempt to observe two times
per night. This would allow them to call alerts relatively early
for Jupiter-mass planets, and to have a better chance of finding
evidence of Earth-mass \pl s. In addition, cooperation among the teams
could greatly enhance the probability of finding reliable evidence of
short events. At a recent workshop on \ml , people associated with three
of the teams
discussed the advantages of choosing one or two fields that all of the
teams would attempt to monitor each night. A motivating factor for
such cooperation is to better understand the relative detection
efficiencies of the teams.  The quest for short \ev s provides another
important motivation. Indeed, if each team visited one or two Bulge
fields twice per night, there would at least occasionally, be
6-8 times per night that those fields were checked. If, on average,
10 \ev s per year were to be discovered in those fields,
then, over the course of 3 years, short \ev s could be discovered or
ruled out at the $\sim 3\%$ level.
In addition, the teams should call alerts for \shdn\ \ev s,
even if the \ev\ has apparently ceased before they can announce
its discovery. If the \shdn\ \ev\ is due
to lensing by a \pl, calling the alert 
will allow the follow-up teams to have a better chance of
detecting any subsequent repeat that might be due to lensing by
another object  
in the \ps .
 
(4) {\sl Use of Pixel Techniques:}\ During the past three years, pixel
techniques have begun to be used for the study of \ml\
in M31 (Tomaney \& Crotts 1996; Crotts \& Tomaney 1996; Crotts 1996;
Ansari {\it et al.} 1997; Han 1996).  It has
been estimated that, were such techniques to be
applied to the LMC and Bulge fields, the rate of event detection
would increase by a factor of $\sim 2-3$ (Crotts 1997; Kaplan 1997). 
The reason for the
increase is that observable \ev s can occur when the 
baseline 
flux we receive from a star is not 
bright enough for the star to appear on the templates
presently used by the teams, if the flux
is brought above the detection limits
through \ml. Such \ev s are presently missed by the \mo\ teams.
This increase in detection efficiency, would be helpful, particularly
in any fields singled out for frequent \mo . Indeed, the MACHO team
is presently engaged in applying pixel subtraction techniques to 
their LMC data in an attempt to discover \ev s that were missed by
their standard methods of detection. Application of these techniques to
the
Bulge fields would be also helpful.

\subsubsection{Useful Modifications: The ``Follow-Up" Teams}

(1) {\sl Repeating Events:}\ The follow-up teams need to continue
to monitor all \ev s after the flux has apparently fallen back to
baseline. Note that this includes resonant lensing  \ev s, since 
(a) if the observed \ev\ was due to a \pl, there may
well be a second planet, and (b) if the observed \ev\ was due to a 
binary lens, one or
both members of the
 binary (or even the combination) may support a
\ps . 

We note that, even when the \mo\ teams have identified a \shdn\ \ev\ that 
appears to have ceased before more frequent \mo\ could begin, it is worthwhile
for the follow-up teams to continue to monitor the flux. If the
short-time-scale \ev\ was due to lensing by a \pl, the
result of frequent \mo\ 
could be the detection of a repetition due to a moon revolving about the \pl.
If a moon exists and if we were fortunate in the orientation of the
source track, then a repetition could occur and should be detected
within a matter of days.
A more delayed repetition could be due to later lensing by the
central star or by another \pl\ in the system.  If they can detect
\ev s with $A_{max}$ as small as $1.06$, the follow-up teams
 will be at least twice as likely to discover such \ev s as are the
\mo\ teams.
 
It should be possible for the follow up teams to
discover repetitions in most cases in which a star is the first lens
encountered. 
For the \ps s we have simulated,
roughly $1/2-3/4$ of all other repetitions could be detected if
monitoring were continued for $100$ days.
In addition, overlap \ev s would not require a significant wait-time,
since the second encounter would start (i.e., be associated with $A > 1.06$)
even  before the first encounter ended.

(2) {\sl Isolated Events:}\
While monitoring known ongoing events, the follow-up teams have many
other stars in their field of view.  When the total number, including those
not individually above the detection limit (but which could be brought above
the limit if magnified by some reasonable amount) is large enough, the
follow-up teams can hope to identify new events.  Such observations would
play a unique role in the identification of new events, particularly the
short-duration events that should be associated with wide \ps s.
Programs that would allow the so-called follow-up teams to take the
lead in \ev\ detection are already underway or are planned 
(Sahu 1997; Gould 1997). Indeed,  
an ideal \ml\ search for the purposes of
the detection of planet lenses, is one in which the follow-up
teams play the role now played by the \mo\ teams to discover
\ev s, and continue the work by \mo\ the \lc s of the \ev s they discover.

\section{Expectations}

What results are likely to be derived if the strategies sketched in
the previous section are utilized? It is difficult to answer this
question, because we know so little
about how common \ps s are, and about
the distributions of planetary masses and
orbital periods. Because, however, planetary systems have begun to be discovered
in the Sun's local neighborhood, it is beginning to seem
likely that many, perhaps most stars support \ps s.

If this is so, then the preliminary results represented by the
known \ps s
provide encouragement that \ml\ will play an important
role in the discovery of \pl s.
The systems listed in Table 1 represent roughly $1/5$
of all of the confirmed planetary and binary brown dwarf systems.
Each would be a good candidate for detection via \ml . 
 PSR B1620-26 and the brown-dwarf system  Gl 229 would
produce larger numbers of isolated \ev s than our solar system, but
fewer repeating \ev s. 
In addition, 55 Cnc and HD 29587 are
excellent candidates for lenses that would lead to resonant
\ev s.
\footnote{47 Uma has an estimated
orbital separation slightly less than 0.8 $R_E$ for
$D = 10 kpc$, $x = 0.9$. It may also be a good candidate for detection.}
We note that, since \pl s in \wo s, and even those in the \zres ,
 are certain to
be under-represented in our present census of \pl s, inferences
based on Table 1 may be conservative.
Thus, our present knowledge of \ps s makes it seem
likely that the observing teams
will observe some events   
over the next few years.

More important than the detection of any individual \ev, however,
is our ability to extract information about the population of \ps s
in the regions surveyed by the \ml\ teams. We would like to
know the answers to basic questions: what fraction of stars have
\ps s? What are typical numbers of \pl s in a single \ps ? 
What are the distributions of \pl\ masses and orbital periods? 

If the teams begin to discover \pl s,  
they will likely be able to 
establish the statistics of and distributions of
properties among \ps s in the Bulge.
\S 10.3 is devoted to exploring this issue
in more detail.

\subsection{Other Applications}

Before continuing with the study of \pl s, it is
worthwhile to consider what the teams using the wide-planet-search strategies
will learn, even if nature has been so unkind as to neglect to
provide most stars with \pl s. Planet-motivated
investigations will yield interesting fruit
regardless of the size of the population of \pl s. 

First, they will increase the detection rates of all \ev s,
particularly \shdn\ \ev s.
Second, the frequent \mo\ with good photometry will
allow us to learn more about each \ev\ detected.
Common astronomical effects, such as stellar binarity,
blending, and \fss ,  
 are expected to significantly
affect the shape of lensing \lc s, 
introduce time dependence into the spectra, and even produce 
apparent repetitions.   
(See e.g, Griest \& Hu 1992; Mao \& Paczy\'nski 1991;
Mao \& \rd\ 1995;    
\rd\ \& Esin 1995; Kamionkowski 1995; 
Loeb \& Sasselov 1995; Simmons, Willis \& Newsam 1995; \rd\ \& Mao 1996;
Sasselov 1997;
\rd\ \& Perna 1997; \rd\ 1997) Studying the
manifestations of these effects in the data sets can
(1) can break the degeneracy of individual \lc s, and 
(2) allow us to learn more about the
populations of sources and lenses.
Along the direction to the Bulge, we will be able to learn 
a good deal about the stellar luminosity
and mass functions, and the binary fraction, as well as the distribution
of binary properties.  
The  information we collect can inform our design of the next
generation of \ml\ observations. For example, satellite projects have been
proposed and preliminary calculations indicate that they are likely   
to be productive. (See, e.g., Boutreux \& Gould 1996 and references therein.)
The detailed observations that would be made as part of the search
for \pl s in \wo s would provide  solid input, useful for the detailed
 planning needed for  
such space-based projects. 

\subsection{The Relative Numbers of  ``Resonant" and ``Wide" Events}

Without knowing more about the distributions of planet properties--
which is exactly
what we are trying to learn about through the proposed \ml\ studies--
it is not possible to make definitive predictions for the relative
numbers of events due to \pl s in the zone for resonant lensing, 
and \ev s due to \pl s in \wo s.
We do know enough, however, to understand the issues that
determine the relative \ev\ rates.

If we place a Jupiter-mass planet in a resonant orbit, the chance of
detecting evidence of the planet's presence is close to $20\%$
The present observing set-up is optimized to discover this type of
\ev . If we place the same \pl\ in a wide orbit, the probability of
detecting an isolated \shdn\ \ev\ due to the \pl\ is smaller,
approximately $(3\, n)\%$ (where $n$ is defined by Eq.\ 4),  
and the probability of detecting a repeating \ev\ in which the central star
serves as the other lens is $(6\, n^2 \%)/(\pi\, a).$ 
Thus, with the present
observing strategy, individual examples of
each type of wide-orbit \ev\ are much less likely to
be observed than a resonant \ev. 
Even with the present set-up, however,
\pl s in \wo s may be detected at a rate comparable to 
or even larger than the rate of detecting \pl s in  
resonant orbits, simply because
on average there may be on the order of $10$ times as many of them as there 
are \pl s in resonant orbits. 

Two factors can enhance the relative 
probability of detecting \pl s in \wo s.
The first is that improved sensitivity to 
\shdn\ \ev s and   
better photometric sensitivity 
can significantly increase the detection
rates for all wide-orbit \ev s, particularly
of repeating \ev s.
The second is the influence of \fsse , which
can also  
effectively increase the  
value of $n$, even for a fixed value of $A_{min}.$

\begin{deluxetable}{lllll}
\scriptsize
\tablecaption{Detection Rates for Planets in Resonant and Wide Orbits:
\hfil\break
Point and Finite-Sized Circular Sources}
\tablehead{\colhead{Detect \tablenotemark{(1)} } &
   \colhead{$P_{res}$ \tablenotemark{(2)} } &
   \colhead{$P_1-P_\odot$ \tablenotemark{(3)} } &
   \colhead{$P_1^{overlap}$ \tablenotemark{(4)} } &
   \colhead{$P_2$ \tablenotemark{(5)} }} 
\startdata
\multicolumn{5}{l}{\bf Solar system, $V =$ Gaussian,
   $\alpha = $ uniform:} \\ \\
A & 5 & 0.3 & 2.1 & 0.7 \\
B & 5 & 1.4  & 4.2 & 1.8  \\
C & 5 & 3.1  & 4.2 & 2.1  \\
\hline
\multicolumn{5}{l}{\bf Solar-like system w/finite source, $V =$ Gaussian,
   $\alpha = $ uniform:} \\ \\
C & 0.3 & 2.4  & 0.7 & 1.0 \\
\hline
\multicolumn{5}{l}{\bf Power-of-3 system, $V =$ Gaussian,
   $\alpha = $ uniform:} \\ \\
A & 10-20 & 13 & 1.9 & 1.5 \\
B & 10-20 & 29 & 3.6 & 3.1 \\
C & 10-20 & 30 & 3.6 & 3.1 \\
\hline
\multicolumn{5}{l}{\bf Power-of-3 system w/finite source, $V =$ Gaussian,
   $\alpha = $ uniform:} \\ \\
C & 0.7-1.3 & 4.9 & 0.7 & 0.5 \\
\enddata
\tablenotetext{(1)}{Descriptions of the detection conditions can be found in
the text.  All probabilities are given as percentages of the number of events
in which the central star
 was encountered by itself, and in which the magnification
reached at least $A_{min} = 1.34$.  
}  
\tablenotetext{(2)}{
We have 
have used the results of Bennett \& Rhie (1996) to compute the probability
that a \pl\ located there will be detected. 
They considered a mass ratio of $10^{-5}$, which, 
corresponds to an Earth-mass \pl\ if the \cs\ has a mass
of $\sim 0.3 M_\odot.$ Thus, when we use the results from
Bennett \$ Rhie (1996) to derive the number of resonant \ev s
expected in our models, 
we may be making a slight overestimate.  
To be in line with $A_{min}$ for the wide-orbit calculations,
we have assumed that the observed resonant-\ev\ perturbation
should be a $6\%$ deviation from the underlying \lc , 
and have interpolated between
the results for the computation for $4\%$ and $10\%$ effects
(Bennett \$ Rhie 1996).
We have also averaged uniformly weighted
contributions for all of the cases they considered,
which correspond to $r\sim 1, 2, 4.3, 10,$ even though for the wide-orbit
case, where we effectively also performed a linear average, we
required $3<r<7.$
For the solar system, we have assumed that there is a $25\%$ chance that one
of the \pl s is in the \zres; 
for the power-of-3 model we have assumed that there
is a $50-100\%$ chance that a \pl\  is in the \zres.}
\tablenotetext{(3)}{Percentage of
isolated  (non-repeating) events (one ``peak'' in
the light curve) with no overlap (see below).}
\tablenotetext{(4)}{Percentage of isolated (non-repeating)
overlap events. In this case the two lenses were the central
star and the innermost \pl .
}
\tablenotetext{(5)}{Percentage of events with one repetition
(two ``peaks''). Overlap \ev s are not included.}
\end{deluxetable}

Table 4 illustrates the situation for our solar system, and for a solar system
composed of Earth-mass \pl s. (See also Figure 15.)
The first three lines pertain to
a \ps\ identical  
to our own solar system, placed in the Bulge (see also Table 3).
Since, averaging over angles, Jupiter has a $20-25\%$ chance of being     
viewed in the \zres, and since there is a $\sim 20\%$ chance of
detecting a Jupiter-mass in the \zres,
we have estimated that
there is a $\sim 5\%$ chance of detecting evidence of the
solar system through a resonant lensing \ev . For the detection
criteria of set A, the resonant-lensing signature
 would be the dominant mode of detection. Changing the
detection strategy to allow $A_{min}=1.06$ ($n=2$), allows wide-orbit
\ev s to dominate, with overlap and repeating \ev s having a combined
detection
rate of $6.3\%$ ($7.3\%$ if we eliminate the requirement of a
$1-$day \du\ for the first \ec ). Thus,
overlap and repeating \ev s are competitive with resonant lensing \ev s,
even if we assume that we may catch
most of the overlap \ev s but only $1/2$ of the repeating \ev s. 
Isolated \pl-lens \ev s would be found at a rate of $1.4\%$ and $3.1\%$
for detection criteria B and C, respectively.  These results indicate
that the
conjecture that all stars have \ps s similar to our solar system,
predicts  
the detection of resonant, repeating, and \shdn\ isolated \ev s in roughly
equal numbers. If, however, all of the \pl s in the model solar system 
were of 
Earth mass, the detection rate would fall. The fall is much more precipitous
for resonant \ev s, which are $5$ times less likely than the combined
rate of repeating and overlap \ev s and $8$ times less likely than
isolated \shdn\ \ev s.

For the power-of-3 model, we have assumed that the probability
of finding a \pl\ in the \zres\ ranges from $50\%$ to $100\%.$
The general pattern
of relative rates is similar
to that for the solar system. For Jupiter-mass \pl s,
resonant events are competitive and
can even be dominant when the detection criteria of set A are used. 
The detection criteria
B and C increase 
the detection rate of overlap and repeating \ev s.
Isolated \shdn\ \ev s can also be important, occurring at a higher than  
resonant \ev s. 
Finite-source-size   
effects decrease the overall detection rates, making the 
repeating and overlap \ev s as common as resonant \ev s, 
and increasing the 
relative importance of isolated \pl-lens \ev s. 

The general pattern of relative rates illustrated by these examples is likely to 
be reflected in our data sets.
That is,  when \fsse\ are not important, the wide-orbit 
discovery channel we
have studied is competitive with the resonant-\ev\ channel. 
Repeating and overlap \ev s tend to occur at a lower rate than
either resonant or isolated \shdn\ \ev s, but the rates of the latter
two types of \ev s are comparable.  
When \fsse\
are important and/or when a strategy to optimize the discovery of \pl s
in \wo s is implemented, discovery of wide-orbit \pl s may dominate.
Isolated \shdn\ \ev s should be the \pl-lens \ev s most frequently detected,
and the combined rate of repeating and overlap \ev s may be 
comparable to or, depending on the influence of finite-source size,
 somewhat
larger than the rate of resonant-zone lensing \ev s.

\subsection{The Populations of Planetary Systems and Low-Mass MACHOs}

\subsubsection{Planetary Systems: General Considerations}

We found in \S 4 that we could hope to test simple hypotheses about specific
types of \ps s, even within the next few years. For example, do most
stars have \ps s similar to the solar system? or similar to the power-of-2
or power-of-3 models?    
Of course it is most likely that \ps s come in several different varieties.
We can use the data to systematically extract information
about lenses with a possibly complicated distribution of \ps\
properties as follows.

Repeating and resonant events each allow us to determine specific
features of the \ps\ that served as a lens. Each provides the value
of the projected separation between one planet and another object
(either the sun for resonant events [and most repeating \ev s],
or a second planet for some repeating \ev s) in the \ps. Each allows
the derivation of a mass ratio.
\footnote{
The extraction of \pl\ parameters, the degeneracies in those
parameters, and the degeneracies between the \pl-lens interpretation and other
effects have been studied for \pl s in the \zres\  
by Gaudi \& Gould (1997) and Gaudi(1997).
Parameter extraction should generally be more straightforward for
\pl-lenses in \wo s. If the lens separations are very large, then the
\lc s are much like the standard point-lens \lc s; if the separations
are smaller, the fit given by \rd\ \& Mao (1996) can be applied.} 
 If more information is available,
derived from evidence of blending or finite-source-size effects, for
example, it can even be possible to place reasonably tight limits
on the mass of the planet-lens.
If we are able, therefore, to discover and analyze a number
of repeating and/or resonant planetary-system-lens \ev s, we will, by
also including the influence of observational selection effects, be
able to derive some of the characteristics of the population
of \ps s in the Galactic Bulge and elsewhere. 

Consider 
a population of \ps s that act as lenses. Let 
${\cal P}_{res}(m)$ be the probability that a \pl\ of mass $m$ is
in the \zres, and ${\cal P}_{w}(m,a)$ be the probability that 
a \pl\ of mass $m$ is in a \wo, with separation $a$ from the central star.
As described above, discovering and measuring the rates
 of repeating and resonant \ev s 
constrains the form of these probability functions.
Furthermore, there is a consistency check, since
the values of ${\cal P}_{res}(m)$ and ${\cal P}_{w}(m,a_w)$
must match.  
The integral, $\int_{a_w}^\infty da\, {\cal P}_{w}(m,a),$
then predicts the rate of isolated \shdn\ \ev s that should be due to
lensing by \ps s.

\subsubsection{Low-Mass MACHOs}

Should the predicted rates of detectable
short-duration \ev s be larger than the numbers actually observed,
then the analysis of \ps\ lenses must be
carefully reconsidered. We expect, however, that if there are
deviations from the predictions, they should be 
because the observed rate is higher than the predicted rate.
This is because \pl s in \ps s are not the only
lenses expected to produce short-time-scale \ev s.
For example, orbital dynamics may lead some \pl s
to be ejected from their home \ps .
If even $10\%$ of massive planets
are so-ejected, then, considering that the \ml\ observations
are sensitive to the planet debris of many generations of stars,
some of these ejected \pl s could be detected. In addition,
it is certainly possible that a significant fraction of Galactic
dark matter 
exists in the form of low-mass objects.
It is an interesting fact that, without the information  provided
by repeating and resonant \ev s, it will not be possible to
determine whether there is a low-mass MACHO component in directions
in which lensing by stellar systems contributes significantly to the
rate of \ml .
Thus, when stars contribute significantly to the rate of lensing,
 the detection of repeating and resonant \ev s is important 
in helping us to 
learn about any Galactic component of low-mass MACHOs.

\section{Summary}

We have introduced the tools needed to systematically use \ml\ to
search for \pl s in wide orbits around distant stars. Because
these searches can be conducted as part of the ongoing \ml\
observations, we have carried out detailed simulations to determine
what signatures should be expected if our solar system or other
known or model systems serve as lenses. We have also studied how
different detection strategies influence detection efficiencies.
Our results are encouraging in that they clearly indicate that
a systematic search for \pl s in \wo s is not only feasible,
but that, even over the short term,
it can yield interesting results about the population
of \ps s in our own and other galaxies. 

Until now, \ml\ searches for \pl s have concentrated on searching
for \pl s that might be located in the zone for resonant lensing.
Resonant lensing \ev s are expected and will play an important role
in our quest to learn more about \pl s through \ml .
Planetary systems, however, seem likely to exhibit enough structure,
in the form of multiple \pl s, moons revolving about \pl s, and even
belts of compact debris, that it is important to explore all the ways
\ml\ can help us to study them. 
We point out that \pl s in wide orbits ($a > 1.5 R_E$)
(1) should exist in larger numbers than planets in the \zres ,
(2) also can yield distinctive signatures, and 
(3) may yield larger numbers of detectable \ev s, particularly
if \fsse\ are important. 
Thus, previous estimates of the \ml\ detection rate (see, e.g., Peale 
1997) need to be
revised upward.

The key elements of detectability are to pick up evidence  
of \ml\ \ev s as early as possible, and to be sensitive to \ev s
 even if the peak 
\mage\ is smaller than $1.34.$ 
The detection rate of isolated \pl-lens \ev s is proportional to $n$,
where $n=w_i/R_{E,i}$, and $w_i$ represents the distance
of closest approach needed for reliable detection of lens $i$.
The rate for repeating events scales as $n^{k+1},$ where $k$ is the 
number of repetitions. Just as is the case for the detection of \pl s 
in the \zres , frequent monitoring of ongoing \ev s is important,
although hourly \mo\ may not be necessary.

The features that will improve the rate of \pl\ detection
should be considered when designing the next generation of
\ml\ observations. 
In the meantime, 
the MACHO results to date indicate that the teams have the 
necessary capabilities to detect wide-orbit \pl s.
 We have suggested a set of modifications
 in the present detection strategy that could, in the short-term,
 significantly
improve detection rates for isolated and repeating \pl-lens \ev s.
One important component is to be alert for the possibility that
repeating \ev s may be \ml\ \ev s, and to continue to
carry out frequent follow-up \mo , even after an
apparently isolated \ev\ has ceased.
It is true that the likelihood of finding a repetition may be 
on the order of a few percent, but the  relative importance
of the \ev s makes the study worthwhile. After all, \ml\ itself
is a low-probability phenomenon, but has been well worth looking for.
In fact, repeating \ev s must be present for a number of 
reasons in addition to the wide-planet connection (Griest \& Hu 1992,
\rd\ \& Mao, 1996).
Nevertheless, almost alone 
among all of the \lc\ perturbations suggested so far, the search for them
has remained something of a taboo.    

An important consequence of the likely presence of \pl s in a stellar population
being studied for signs of \ml, is that isolated short \ev s are 
very likely to be present at a level that can be as high (compared to
single stellar-lens \ev s) as ${\cal O}(10\%)$. 
This means that any signature due to low-mass MACHOs cannot be 
unambiguously identified unless the contribution due to \pl s can first be
quantified. (There are exceptions when, for example,
it is known that the majority
of lensing \ev s cannot be due to ordinary stellar systems,
or when the rate of short-time-scale \ev s is so high that the associated
optical depth is larger than could possibly be due to \ps s.) 
Fortunately, studies of repeating and resonant \ev s, and any
short-duration \ev s subject to either \fsse\
or the blending of light from the \cs\ with that of light from the lensed star, 
should allow the contribution of \pl s to \shdn, 
apparently isolated \ev s to be quantified.
This contribution can then be subtracted from the total
to derive the magnitude of any contribution due to dark matter existing
outside the realm of 
ordinary stellar systems.   

Another point we have emphasized is that, even though the \ps s
\ml\ discovers are far away, some may nevertheless be
the subjects of fruitful further study. Indeed, when the central
star is luminous, we can hope to determine its spectral type.
In some cases, this can help to set the mass scale for the system, and
can therefore
help us to determine the mass of the \pl s that served as lenses.
Finite-source-size effects can also put constraints on the
lens mass.
Thus, although we will not 
image beach front property on the 
\pl s discovered via \ml , 
we should not give up on the possibility of learning more
about individual \ps s that serve as microlenses.
We have also pointed out that it is precisely in those systems
in which a \pl\ discovered via \ml\ is most
likely to have Earth-like conditions, that the central star may be
luminous enough to permit further study.

Searches for planets using \ml\ should be able to extend the reach of
local planetary searches by discovering planets in distant parts
of our own and other galaxies and by discovering even low-mass \pl s
orbiting at low speeds. The search for planets in
wide orbits
represents a significant extension of the ongoing microlensing searches.
Indeed, it seems likely that \pl s in \wo s will
provide an important, and
possibly even the dominant mode for the detection of planetary systems
via microlensing, particularly Earth-mass \pl s.

\bigskip
\bigskip
\bigskip
\centerline{ACKNOWLEDGEMENTS}

One of us (RD) would like to thank 
Arlin Crotts, Andrew Gould, Jean Kaplan, Christopher Kochanek, 
David W. Latham, Avi Loeb,   
Shude Mao, Robert W. Noyes, Bodhan Paczy\'nski,
Bill Press, Michael M. Shara, Edwin L. Turner,  Michael S. Turner,  
and the participants in the 1997 Aspen workshops, ``The Formation
and Evolution of Planets" and ``Microlensing"
for interesting
discussions, and the Aspen Center for Physics and the Institute for
Theoretical Physics at Santa Barbarba for their hospitality while
this paper was being written.
 One of us (RAS) would like to thank the 1996 CfA Summer
Intern Program for support and the Harvard-Smithsonian Center for
Astrophysics for its hospitality while the work was underway.
This work was supported in part by NSF under GER-9450087 and
AST-9619516.

{}
\end{document}